\documentstyle[11pt]{article}
\newcommand{\CMP}[1]{{\em Commun. Math. Phys.} {\bf {#1}}}
\newcommand{\JMP}[1]{{\em J.~Math. Phys.} {\bf {#1}}}

\newcommand{\HPA}[1]{{\em Helv. Phys. Acta} {\bf {#1}}}
\newcommand{\PL}[1]{{\em Phys. Lett.} {\bf {#1}}}

\newcommand{\Ann}[1]{{\em Ann. Phys.} {\bf {#1}}}

\newcommand{\restr}[1]{\left.\right|_{#1}}

\newcommand{\map}{\mbox{\em map}\,}
\newcommand{\Map}{\mbox{\em Map}\,}

\newcommand{\eq}{\begin{equation}}
\newcommand{\eqend}{\end{equation}}
\newcommand{\eqa}{\begin{eqnarray}}
\newcommand{\nonueqa}{\begin{eqnarray*}}
\newcommand{\eqaend}{\end{eqnarray}}
\newcommand{\nonueqaend}{\end{eqnarray*}}
\newcommand{\nonu}{\nonumber \\ \nopagebreak}
\newcommand{\bma}[1]{\begin{array}{#1}}
\newcommand{\ema}{\end{array}}
\newcommand{\bc}{\begin{center}}
\newcommand{\ec}{\end{center}}

\newcommand{\Ref}[1]{(\ref{#1})}

\newcommand{\lrar}{\longrightarrow}

\newcommand{\dd}{\mbox{${\rm d}$}}

\newcommand{\ee}[1]{\mbox{{\rm e}}^{#1}}
\newcommand{\ii}{{\rm i}}
\newcommand{\OO}{{\rm O}}
\newcommand{\sgn}[1]{\mbox{{\rm sign}}({#1})}

\renewcommand{\det}[1]{\mbox{{\rm det}} ({#1})}
\newcommand{\detF}[1]{\mbox{{\rm det}}' ({#1})}

\newcommand{\detz}[1]{\mbox{{\rm det}}_2 ({#1})}
\newcommand{\detp}[1]{\mbox{{\rm det}}_p ({#1})}
\newcommand{\detzp}[1]{\mbox{{\rm det}}_{2p} ({#1})}

\renewcommand{\vec}[1]{\mbox{\boldmath ${#1}$}}

\newcommand{\Gam}{\Gamma}

\renewcommand{\phi}{\varphi}
\newcommand{\sig}{\sigma}
\newcommand{\del}{\delta}

\newcommand{\Om}{\Omega}

\newcommand{\eps}{\varepsilon}
\newcommand{\R}{{\sf I} \! {\sf R}}
\newcommand{\C}{{\sf C} \! \! \! {\sf I}\:}
\newcommand{\Cb}{{\sf C} \! \! \! {\sf I}\:^\times}
\newcommand{\N}{{\sf I} \! {\sf N}}

\newcommand{\f}{\frac}

\newcommand{\cA}{{\cal A}}

\newcommand{\cD}{{\cal D}}
\newcommand{\cL}{{\cal L}}

\newcommand{\cH}{{\cal H}}
\newcommand{\cF}{{\cal F}}
\newcommand{\cE}{{\cal E}}
\newcommand{\cG}{{\cal G}}
\newcommand{\cU}{{\cal U}}
\newcommand{\cS}{{\cal S}}

\newcommand{\FGam}[1]{\Gamma ({#1})}                          
\newcommand{\dFgam}[1]{J({#1})}                 
\newcommand{\hGam}[1]{\hat{\Gamma}  ({#1})}          
\newcommand{\dhgam}[1]{\hat{J} ({#1})}  
\newcommand{\hGamp}[1]{\hat{\Gamma}'  ({#1})}          
\newcommand{\hzGam}[1]{\hat{\Gamma}_{2}  ({#1})}     
\newcommand{\hpGam}[1]{\hat{\Gamma}_{p}  ({#1})}     
\newcommand{\dhpgam}[1]{\hat{J}_p({#1})}  
\newcommand{\ccr}[2]{{[} {#1},{#2} {]} }        
\newcommand{\CCR}[2]{{[\!\![} {#1},{#2} {]\!\!]} }        
\newcommand{\car}[2]{ \{ {#1},{#2} \}  }        
\newcommand{\produ}[2]{( {#1},{#2} ) }          
\newcommand{\Produ}[2]{< {#1},{#2} > }          

\newcommand{\tra}[1]{{\rm tr} ({#1})}          
\newcommand{\norm}[1]{|{#1}| }                 
\newcommand{\Norm}[1]{|\! | {#1} |\! |}        
\newcommand{\NORM}[1]{|\! |\! |{#1} |\! |\! |} 

\newcommand{\I}[1]{B_{#1} (h)}     
\newcommand{\ghe}[1]{\vec{g}_{#1}(h;\eps)}          
\newcommand{\Ghe}[1]{\vec{G}_{#1}(h;\eps)}          
\newcommand{\wghe}[1]{\widehat{\vec{g}_{#1}}(h;\eps)}
\newcommand{\wGhe}[1]{\widehat{\vec{G}_{#1}}(h;\eps)}

\newcommand{\Grhe}[1]{Gr_{#1}(h;\eps)}

\newcommand{\ghF}[1]{\vec{g}_{#1}(h;F)}          
\newcommand{\GhF}[1]{\vec{G}_{#1}(h;F)}          

\newcommand{\Che}[1]{\vec{C}_{#1}(h;\eps)}
\newcommand{\che}[1]{\vec{c}_{#1}(h;\eps)}
\newcommand{\Uhe}{\cU(h;\eps)}
\newcommand{\UhF}{\cU(h;F)}
\newcommand{\UheF}{\cU^{(F)}(h;\eps)}
\newcommand{\CChe}[1]{\vec{C}_{#1}^{c}(h;\eps)}

\newcommand{\She}[1]{\vec{\cS}_{#1}(h;\eps)}

\newcommand{\mult}{\star}

\newcommand{\multz}{\hat{\star}_2}
\newcommand{\multp}{\hat{\star}_p}

\newcommand{\remark}{\subsubsection*{Remark:}}

\newcounter{saveeqn}
\newcounter{App} 
\newcommand{\app}{%
\stepcounter{App}%
\setcounter{saveeqn}{\value{equation}}%
\setcounter{equation}{0}%
\renewcommand{\theequation}{\Alph{App}\arabic{equation}} }
\newcommand{\appende}{%
\setcounter{equation}{\value{saveeqn}}%
\renewcommand{\theequation}{\arabic{equation}}  }
\newcommand{\U}{{\rm U}}
\newcommand{\SU}{{\rm SU}}
\renewcommand{\subset}{\subseteq}

\author{Edwin Langmann\thanks{Erwin Schr\"odinger-fellow,
supported by the ``Fonds zur F\"orderung
der wissenschaftlichen Forschung'' under the contract Nr. J0789-PHY}
\\ Department of Physics\\ The University of British Columbia\\
V6T 1Z1 Vancouver, B.C., Canada}
\title{Fermion Current Algebras and Schwinger Terms in
$(3+1)$-Dimensions}

\begin{document}
\thispagestyle{empty}
\maketitle
\vspace*{30.0mm}
\thispagestyle{empty}
\begin{abstract}
We discuss the restricted linear group in infinite dimensions modeled
by the Schatten class of rank $2p=4$ which contains
$(3+1)$-dimensional analog of the loop groups and is closely related to
Yang-Mills theory with fermions in $(3+1)$-dimensions. We give an
alternative to the construction of the ``highest weight'' representation
of this group found by Mickelsson and Rajeev. Our approach is close to
quantum field theory, with the elements of this group regarded as
Bogoliubov transformations for fermions in an external Yang-Mills
field. Though these cannot be unitarily implemented in the physically
relevant representation of the fermion field algebra, we argue that
they can be implemented by sesquilinear forms, and that there is a
(regularized) product of forms providing an appropriate group
structure. On the Lie algebra level, this gives an explicit,
non-perturbative construction of fermion current algebras in $(3+1)$
space-time dimensions which explicitly shows that the ``wave function
renormalization'' required for a consistent definition of the currents
and their Lie bracket naturally leads to the Schwinger term identical
with the Mickelsson-Rajeev cocycle.  Though the explicit form of the
Schwinger term is given only for the case $p=2$, our arguments apply
also to the restricted linear groups modeled by Schatten classes of
rank $2p=6,8,\ldots$ corresponding to current algebras in $(d+1)$-
dimensions, $d=5,7,\ldots$.
\end{abstract}
\section{Introduction}
\label{sec1}

In recent years it has become obvious that the representation theory
of certain infinite dimensional Lie groups and algebras can contribute
in an essential way to the understanding of quantum field theory.
Well-known examples are the Virasoro algebra and the affine Kac-Moody
algebras which have been of crucial importance for recent spectacular
progress in two dimensional conformal field theory.

The groups $\Map(M^d;G)$ of maps from a $d$-dimensional manifold $M^d$
to some (compact, semisimple) Lie group $G$ naturally arise as gauge
groups for Yang-Mills theories on space-time $M^d\times \R$ in the
Hamiltonian formalism. General principles of quantum theory imply that
the physical Hilbert space of such models should carry a
highest-weight representation of the gauge group. This strongly
suggests that such representations of $\Map(M^d;G)$ and its Lie
algebra $\map(M^d;g)$ ($g$ the Lie algebra of $G$) should be of
crucial importance for quantum gauge theories, and, on the other hand,
that quantum field theory should give a natural guide for finding
interesting representations of these groups.

Indeed, the by-now well-understood representation theory of the loop
groups $\Map(S^1;G)$ and loop algebras $\map(S^1;g)$ \cite{PS,KR} has
provided a general, rigorous basis for $(1+1)$-dimensional quantum
gauge theories. Especially, the wedge representation of $\Map(M^d;G)$
and $\map(S^1;g)$ allows for a complete understanding and accounting
for all ultra-violet divergences arising in the fermion sector of
dimensional Yang-Mills theories with fermions\footnote{In this paper I
am discussing only quantum field theories with fermions.  I am
planning to report on corresponding results for bosons in a
forthcoming publication.} on a cylinder (= space-time $S^1\times\R$)
\cite{M3}. The chiral (left- and/or right-handed) fermion currents
arising naturally in such models with massless fermions, are
identical with the ones proving the standard wedge representation of
the affine Kac-Moody algebra $\widehat{\map(S^1;g)}$ (= central
extension of $\map(S^1;g)$), and the cocycle giving the central
extension has a natural interpretation as Schwinger term.  Moreover,
it is such Schwinger terms which are responsible for anomalies: For
gauge theories with Weyl (=chiral) fermions they lead to the
gauge anomaly in the commutators of Gauss' law generators, though
there is no such gauge anomaly for Dirac (=non-chiral) fermions (the
Gauss' law generators involve a sum of the left- and right handed
chiral currents leading to Schwinger term with opposite signs which
can be arranged such that they cancel), these Schwinger terms show up
in the commutators of the temporal and the spatial components of the
vector current and lead to the chiral anomaly \cite{Jackiw,LS2}.

A central role in these developments has been played by the infinite
dimensional group $\vec{G}_1$ modeled by the Hilbert-Schmidt class and
for which a well-developed representation theory exists \cite{PS,M3}.
This group contains all loop groups as subgroups, and representations
of the latter are naturally obtained by restriction from the ones of
$\vec{G}_1$ \cite{PS}. The interesting representations of
$\vec{G}_1$ are not true but projective ones and correspond to true
representations of a central extensions $\widehat{\vec{G}_1}$ of
$\vec{G}_1$.  The most important of these is the wedge representation
which is a highest weight representation of $\widehat{\vec{G}_1}$ on
the fermion Fock space \cite{M3}.

{}From a quantum field theory point of view, $\vec{G}_1$ is identical
with the group all Bogoliubov transformations which are unitarily
implementable in the physical Hilbert space of some arbitrary model
of relativistic fermion in an external field, and the construction of
the wedge representation of $\widehat{\vec{G}_1}$ is equivalent to
constructing the implementers of these Bogoliubov transformations
\cite{BT4,R1,M3}.  This requires some regularization which, on the Lie
algebra level, corresponds to normal ordering of fermion bilinears and
which can be done with mathematical rigor \cite{BT4}. The very
definitions of $\vec{G}_1$ and its Lie algebra $\vec{g}_1$ can be
regarded as a general characterization of the degree of divergence
where this kind of regularization is sufficient. It is this
regularization which leads to the non-trivial 2-cocycle providing the
central extension
\cite{EL3}. On the Lie algebra level one obtains a highest weight
representation of a central extension $\widehat{\vec{g}_1}$ of
$\vec{g}_1$. We refer to the latter as {\em abstract current algebra
in $(1+1)$-dimensions} as it contains all currents referred to
above and also all other operators of interest for gauge theories with
fermions on a cylinder.  This construction also shows on a very
general, abstract level how regularizations can lead to Schwinger
terms implying anomalies \cite{BT4,GL}.

There are two approaches to these current algebras in
$(1+1)$-dimensions: the original one based on the theory of quasi-free
representations of fermion field algebras was pioneered by Lundberg
\cite{Lund} and worked out in all mathematical detail by Carey and
Ruijsenaars \cite{BT4}.  It is conceptually very close to quantum
field theory.  The other approach is by means of determinant bundles
over infinite dimensional Grassmannians \cite{PS,M3} and seems to be
preferred by mathematicians.

In higher dimensions, the situation is much more difficult. This can
be traced back to that fact that the gauge groups $\Map(M^d;G)$ for
$d=3,5,\ldots$ are {\em not} contained in $\vec{G}_1$ in any natural
way --- ultra-violet divergencies are worse in higher dimensions.
There is, however, an infinite dimensional group $\vec{G}_p$,
$p=(d+1)/2$, modeled on a Schatten class and containing $\Map(M^d;G)$
as subgroups \cite{PS,MR}. These groups are the natural starting point
for a generalization of the theory of loop groups and affine Kac-Moody
algebras to higher dimensions. Similar to $(1+1)$-dimensions, it is
natural to regard the definition of $\vec{G}_p$ as a general
classification of the degree of ultra-violet divergencies one
encounters in the fermion sector of gauge theories with fermions in
$(d+1)$-dimensions.

A generalization of the ``Grassmannian approach'' \cite{PS} to the
abstract current algebras from $(1+1)$- to $(3+1)$- and higher even
dimensional space time was developed by Mickelsson and Rajeev
\cite{MR} and Mickelsson \cite{M1,M2} (see also \cite{M3}).  They
were able to construct a ``highest weight''
representation\footnote{the quotation marks indicate that it is not a
highest weight representation in the standard terminology but
something similar \cite{M1}} of an {\em Abelian} extension
$\widehat{\vec{G}_p}$ of $\vec{G}_p$ for $p>1$ on a bundle of fermion
Fock spaces. On the Lie algebra level this leads to a fermion current
algebra with a cocycle corresponding to an operator-valued Schwinger
term depending on a variable in a set $Gr_p$ carrying a non-trivial
representation of $\vec{G}_p$.  Moreover, they tried --- without
success --- to find a Hilbert space $\cH$ such that the currents are
selfadjoint generators of one-parameter families of unitary operators
on $\cH$. Later on it was shown by Mickelsson \cite{M1} and Pickrell
\cite{P} that this is impossible: in higher dimensions than $(1+1)$,
the abstract current algebras of Mickelsson and Rajeev \cite{MR} do
not allow for a faithful, unitary representation on a separable
Hilbert space.

{}From the physical point of view these Abelian extensions of
$\vec{G}_p$ have a natural interpretation: restricting $\vec{g}_p$ to
$\map(M^d;g)$, the fermion currents of Mickelsson and Rajeev
\cite{MR,M1,M2} can be regarded as generators of the gauge
transformations in the fermion sector of Yang-Mills theory with Weyl
fermions. The corresponding restriction of $Gr_p$ can be naturally
identified with the space $\cA$ of all static Yang-Mills field
configuration. The dependence of the Schwinger term on the Yang-Mills
field reflects the fact that in higher dimensions it is not possible
to regularize fermion bilinears independent of the Yang-Mills field.
Indeed, it has been argued already some time ago by Faddeev
\cite{F} and Faddeev and Shatiashvili \cite{FS} that in
$(3+1)$-dimensions such an operator-valued Schwinger term should be
present. It is natural to expect (though to our knowledge it has not
yet been proved) that the Schwinger term of Mickelsson and Rajeev for
$\vec{g}_2$ \cite{MR} when restricted to $\map(M^3;g)$ is identical
with the one of Faddev and Shatiashvili \cite{F,FS}.

In this paper we give an alternative construction of ``highest
weight'' representations of the Abelian extensions
$\widehat{\vec{g}_p}$ of $\vec{g}_p$ for $p>1$ \cite{MR} by means of
the theory of quasi-free representations of fermion field algebras
\cite{BT4}, thus generalizing the construction of abstract current
algebras by Lundberg \cite{Lund} from $(1+1)$- to higher even
dimensional space-time.

We hope that our approach is more transparent than the original one
(at least for physicists); moreover, it shows very explicitly ``how
the Hilbert {space} is lost'' in higher dimensions i.e.\
for the Lie groups $\vec{G}_p$ with $p>1$: as mentioned above,
$\vec{G}_1$ is identical with the group of the Bogoliubov
transformations which are unitarily implementable in the physically
relevant quasi-free representations of the fermion field operators
\cite{R1}, and the implementers with the usual operator
product as group multiplication provide a representation of a central
extension of $\vec{G}_1$ \cite{PS}.  For $p>1$, the elements in
$\vec{G}_p$ correspond to Bogoliubov transformations which are {\em
not} unitarily implementable.  However, it was shown by Ruijsenaars
\cite{R2,R4} that they can be implemented by sesquilinear forms
\cite{Kato}, and these forms can be constructed from the implementers
of the Bogoliubov transformations in $\vec{G}_1$ by some
multiplicative regularization. The operator product of two such forms
does not exist in general, and therefor, in order to get a group
structure for these implementers, one has to regularize also the group
multiplication. The problem is that the usual operator product does
not allow for a multiplicative regularization maintaining
associativity. We show that one can define another product, $\mult$,
for the implementers of the Bogoliubov transformations in $\vec{G}_1$,
and this naturally leads to the group structure of an Abelian
extension of $\vec{G}_1$. We demonstrate that $\mult$ is (not
identical but) equivalent to the operator product, and the resulting
Abelian extension of $\vec{G}_1$ is trivial insofar as it is
equivalent to the central one obtained with the operator product. We
then argue that for each $p\in\N$, there is a unique `minimal'
regularization of the $\mult$-product maintaining associativity, and
this provides a group structure for the regularized implementers
corresponding to the Abelian extension of $\vec{G}_p$ introduced in
\cite{MR} (in fact, we are able to show this explicitly only for
$p=2$, however, the results of \cite{MR,FT} indicate that this is true
for all $p\in\N$).  Moreover, for each $p\geq 1$, the (regularized)
implementers together with the (unique regularized) $\mult$-product
can be naturally interpreted as a representation of
$\widehat{\vec{G}_p}$ on a space of sections in a Fock space bundle
which has the mathematical structure of a module \cite{H} over a ring
of functionals.  For $p=1$, the Fock spaces above different points of
the base space of this bundle are all unitarily equivalent and can be
naturally identified.  However, this is not true for $p>1$.  On the
Lie algebra level, this implies that the construction of abstract
current algebras in $2p$- dimensions, $p>1$, requires not only a
regularization the currents, but one has to regularize their
commutators as well.  Our results show that there is no proper
regularization of the usual commutator of the currents maintaining the
Jakobi identity, but there is (an essentially unique) one for another,
non-trivial Lie bracket of the currents which arises from the
$\mult$-product.

The plan of the paper is as follows. Section \ref{sec2} is
preliminary: we define our notation and motivate the groups
$\vec{G}_p$ and their Lie algebras $\vec{g}_p$ by discussing Weyl
fermions in external Yang-Mills fields, and (for the convenience of
the reader) summarize the basic facts about generalized determinants
which are essential for our construction.  The definitions of the
Abelian extensions $\widehat{\vec{G}_p}$ and $\widehat{\vec{g}_p}$
introduced in \cite{MR}, together with the related Lie group and Lie
algebra cohomology of $\vec{G}_p$ and $\vec{g}_p$ are summarized in
Section \ref{sec3}. A review of the formalism of quasi-free
representations of fermion field algebras is given in Section
\ref{sec4}. The $\mult$-product for the unitary implementers of
Bogoliubov transformations, and the the corresponding Lie bracket for
the abstract current algebra in $2$-dimensions are introduced and
discussed in Section \ref{sec5} and \ref{sec6}, respectively.  In
Section \ref{sec7} we show how the representations of
$\widehat{\vec{G}_p}$, $p>1$ are obtained by an essentially unique
multiplicative regularization, and we derive from this the abstract
current algebras in $(d+1)$-dimensions, $d=(2p-1)$, in Section
\ref{sec8}. We end with a few comments in Section \ref{sec9}.
Details of the calculations are left to five Appendices.

A short account on the results of this paper appeared in
\cite{EL1,EL2} and were discussed in \cite{M5a}.

\section{1-Particle Formalism}
\label{sec2}
\subsection{Fermions in external Yang-Mills Fields}
We consider Weyl fermions in $(d+1)$-dimensional space-time, coupled
to an external Yang-Mills field $A$ with the structure group $G=\U(N)$
or $\SU(N)$ in the fundamental representation.  The space $M^d$ is
assumed to be a smooth (i.e.\ $C^\infty$), oriented, riemannian spin
manifold with a given spin structure, and it is convenient to assume
that the space $M^d$ is compact. (In the case of a non-compact space
(e.g.\ $M^d=\R^d$), we restrict ourselves to Yang-Mills fields and
gauge transformations with compact support).  For $x^j$ some local
coordinates on $M^d$ and $\dd =\sum_{j=1}^d
\partial_j\dd x^j$ the usual exterior derivative, the Yang-Mills field
is a $g$-valued 1-form on $M^d$: $A=\sum_{j=1}^d A_j\dd x^j$, with $g$
the Lie algebra of $G$.

The quantum description of {\em one} fermion can be given in the
Hilbert space $h=L^2(M^d)\otimes V$ of square integrable
functions with values in the finite dimensional Hilbert space $V$
which carries representations of the spin structure and of $G$
and $g$ (to simplify notation, we do not distinguish the
elements in $G$ and $g$ from their representatives on $V$).  On
this 1-particle level, the time evolution is generated by the
usual Weyl Hamiltonian $D_A$ \cite{M3} which is a self-adjoint
operator on $h$, and gauge transformations are given by smooth,
$G$-valued functions $U$ on the space $M^d$:
\eq
D_A\lrar U^{-1}D_A U = D_{A^U}
\eqend
with $A^U = U^{-1}AU + U^{-1}\dd U$ the gauge transformed Yang-Mills
field as usual. We will be mainly concerned with gauge transformations
of the form $U=\exp{(\ii tu)}$ with $t\in\R$, and $u$ some smooth,
$g$-valued function on $M^d$; we denote such an $u$ as {\em
infinitesimal gauge transformation}.  Obviously, every gauge
transformation can be identified with a unitary operator on $h$ (which
we denote by the same symbol), and the group multiplication in the
group $\Map (M^d;G)$ of all gauge transformations is identical with
the product as operators on $h$.  Similarly, the infinitesimal gauge
transformations $u$ can be identified with self-adjoint, bounded
operators $u$ on $h$, and the Lie bracket in the Lie algebra $\map
(M^d;g)$ of all infinitesimal gauge transformations (=pointwise
commutator) is identical with the commutator as operators on
$h$.\footnote{ Strictly speaking, the Lie bracket is $\ii^{-1}
\times\mbox{commutator}$. We find it convenient, however, to
always omit the factor $\ii^{-1}$.  }

\subsection{Linear Groups and Lie Algebras Modelled on Schatten Classes}
\label{sec2.2}
For $p\in\N$, we denote as $\I{p}$ the (so-called) Schatten class
of all bounded operators $a$ on $h$ with a finite norm
\eq
\Norm{a}_p\equiv [\tra{(a^*a)^{p/2}}]^{1/p}
\eqend
($\tra{\cdot}$ is the trace in $h$). Thus $\tra{a^r}$ can be defined
only if $a\in\I{p}$ with $p\leq r$. Especially, $\I{1}$ and $\I{2}$
are the trace class and the Hilbert-Schmidt class, respectively.
By definition, $\I{\infty}$ is the set of all compact operators on
$h$, and $\Norm{\cdot}_\infty=\Norm{\cdot}$ (=operator norm).
Note that $a\in\I{p},b\in\I{q}$ implies
$ab\in\I{r}$ for all $r\geq(\f{1}{p}+\f{1}{q})^{-1}$,
especially $a^p\in\I{1}$
for $a\in\I{p}$.

The essential ingredient for constructing the appropriate
multiparticle theory is $\eps=\sgn{D_A}$ (with $\sgn{x}=1(-1)$
for $x\geq 0(x<0)$ and $\eps$ defined via the spectral theorem for
self-adjoint operators \cite{RS1}) which is a grading
operator\footnote{i.e.\ ,
$\eps^2 = 1$ (identity) and $\eps^*=\eps$ where $*$ denotes the
Hilbert space adjoint } on $h$. Physically, $\eps$ characterizes the
splitting of the 1-particle Hilbert space $h$ in the subspaces
$h_{\pm}=\f{1}{2}(1\pm \eps)h$ of positive ($+$) and negative
($-$) energy states, and it determines the appropriate quasi-free
representation of the fermion field algebra corresponding to
``filling up the Dirac sea'' (see Section \ref{sec4}).

It can be shown that gauge transformations $U\in\Map (M^d;G)$ have
the following crucial property \cite{MR}:
\eq
\label{SI}
\ccr{U}{\eps}\in \I{2p},
\eqend
and the rank $2p$ of the Schatten class is determined by the dimension
of $M^d$:\footnote{The dimension comes in as follows: by direct
calculation one can estimate \protect{\cite{MR}}
\protect{\[
\tra{\ccr{U}{\eps}^*\ccr{U}{\eps}}^p \leq
const.\int_{c}^{\infty}\dd kk^{d-1}k^{-2p}
\]}
($const.$ always finite for smooth functions $U:M^d\mapsto G$) for
some positive $c$, the $\dd k k^{d-1}$ resulting from the volume element
$\dd^dk$.}
\eq
2p = (d+1) .
\eqend
This allows us to embed $\Map (M^d;G)$ in the larger group
$\Ghe{p}$ of all {unitary} operators $U$ on $h$ obeying the
condition \Ref{SI}.  Similarly, $\map (M^d;g)$ can be embedded in
the Lie algebra $\ghe{p}$ of all bounded, self-adjoint operators
$U$ on $h$ obeying this condition
\Ref{SI}. These definitions naturally extend to $p=\infty$.
Note that for all $p\in\N\cup\{\infty\}$,
$\Ghe{p}$ and $\ghe{p}$ both are Banach algebras
with the norm $\NORM{\cdot}_p$ given by
\eq
\NORM{U}_p=\f{1}{2}\Norm{\car{U}{\eps}}+\f{1}{2}\Norm{\ccr{U}{\eps}}_{2p}
\eqend
($\car{\cdot}{\cdot}$ is the anticommutator as usual), and
$\ghe{p}$ is the Lie algebra of $\Ghe{p}$.  In addition, we
introduce the group $\Ghe{0}$ of all unitary operators $U$ on $h$
with $(U-1)\in\I{1}$, and its Lie algebra $\ghe{0}$ containing
all self-adjoint operators in $\I{1}$.

Every gauge transformed Yang-Mills field $A^U$ gives rise to
another grading operator $F=\sgn{D_{A^U}}=U^{-1}\eps U$.  This
suggests to introduce the set $\Grhe{p}$ of all grading operators
of the form $F=T^{-1} \eps T$ for some $T\in\Ghe{p}$, and due to
\Ref{SI}
\eq
(F-\eps)\in\I{2p}
\eqend
$\forall F\in\Grhe{p}$ (it is also easy to see that every grading
operator satisfying this relation is in $\Grhe{p}$).  This set carries
a natural representation $F\to F^U$ of the group $\Ghe{p}$ with
\eq
F^U\equiv U^{-1}FU,
\eqend
and it is a metric space with the metric
\eq
d_p(F,F') \equiv \Norm{F-F'}_p
\eqend
$\forall U\in\Ghe{p}$, $F,F'\in\Grhe{p}$. Note that $\GhF{p} = \Ghe{p}$
and $\ghF{p}=\ghe{p}$ for all $F\in\Grhe{p}$.\footnote{
the norms $\NORM{\cdot}_p$ defined above obviously {\em depend}
on $F$; however, it can be easily shown that
they are equivalent (give rise to the same topology)
for all $F\in\Grhe{p}$
}
Moreover,
\eq
\Ghe{0}\subset\Ghe{1}\subset\Ghe{2}\subset\cdots\subset\Ghe{\infty}
\eqend
and $\Ghe{p}$ is dense in $\Ghe{q}$ for $p<q$,
and similarly for $\ghe{p}$ and $\Grhe{p}$.

\subsection{Generalized Determinants \protect{\cite{S}}}
\label{sec2.3}
For $a$ a linear operator on a Hilbert space $h$, the (Fredholm)
determinant
$\det{a}$ exists if and only if $(a-1)\in\I{1}$, and
$\det{ab}=\det{a}\det{b}$ for all $(a-1),(b-1)\in\I{1}$.  For
$\Norm{a-1}<1$ we have then
\[
\det{a}=\exp{(\tra{\log{a}})}=
\exp{\left\{\sum_{j=1}^{\infty}(-)^{j-1}\f{\tra{(a-1)^{j}}}{j}\right\} },
\]
thus suggesting to define the generalized determinant $\detp{a}$
for $(a-1)\in\I{p}$, $p>1$, by just omitting in this expression the
non-existing traces:
\[
\detp{a}=
\exp{\left\{\sum_{j=p}^{\infty}(-)^{j-1}\f{\tra{(a-1)^{j}}}{j}\right\} }.
\]
Indeed, one can show that for $(a-1)\in\I{p}$,
the operator
\eq
R_p(a)\equiv -1+ a\exp{\left\{\sum_{j=1}^{p-1}(-)^{j}\f{(a-1)^j}{j} \right\}}
\eqend
is in $\I{1}$, hence
\eq
\detp{a}\equiv \det{1+R_p(a)}
\eqend
exists, and for $\Norm{a-1}<1$ it coincides with the expression
given above. Moreover, the mapping $\I{p}\ni a\mapsto \detp{a}$
is continuous in the norm $\Norm{\cdot}_p$, and $a\in\I{p}$ is
invertible if and only if $\detp{a}\neq 0$.

It is natural to regard $\detp{\cdot }$
as the {\em minimal multiplicative regularization} of
$\det{\cdot }$ appropriate for $\I{p}$.
Note that $\mbox{{\rm det}}_1 (\cdot)=\det{\cdot}$, and
$\detp{ab}\neq \detp{a}\detp{b}$
for $p>1$.

\section{Cohomology and Abelian Extensions}
\label{sec3}
In this Section we introduce some terminology needed in the sequel.
\subsection{Abelian Extensions of Lie Groups}
\label{sec3.1}
Let $p\in\N_0\cup\{\infty\}$.  We introduce the set
\eq
\label{b.1}
\Che{p}\equiv \Map(\Grhe{p};\U(1))
\eqend
of `smooth'\footnote{the quotation marks in `smooth' (here and
elsewhere in the paper) are to indicate that we do not give a
precise definition of the meaning of this term and use it only in
a sloppy way as we do not really need all its implications and want to
avoid irrelevant technical discussions. In fact,
all we require is that the differentiations necessary to go
from the group to the Lie algebra level (see the next Section)
are well-defined} functions $\mu:\Grhe{p}\to \U(1); F\mapsto
\mu(F)$; $\Che{p}$ is an Abelian group under point-wise
multiplication:
\eq
\label{b.2}
(\mu\nu)(F)=\mu(F)\nu(F),
\eqend
and it carries a natural representation $\mu\mapsto \mu^U$ of the
group $\Ghe{p}$ with
\eq
\label{b.3}
(\mu^U)(F)\equiv \mu(F^U)=\mu(U^{-1}FU)
\eqend
$\forall\mu\in\Che{p}$, $U\in\Ghe{p}$, $F\in\Grhe{p}$.

`Smooth', $\Che{p}$-valued functions $\beta$ and $\chi$ on
$\Ghe{p}$ and $\Ghe{p}\times\Ghe{p}$ satisfying
\eq
\label{111}
\beta(U^*) = \overline{\beta^{U^*}(U)}, \quad \beta(1)=1
\eqend
(the bar denotes complex conjugation) and
\eq
\label{222}
\chi(V^*,U^*) =
\overline{\chi^{V^*U^*}(U,V)},\quad \chi(U,U^*)=1
\eqend
$\forall U,V\in\Ghe{p}$ are called {\em 1-cochains} and {\em
2-cochains} of the group $\Ghe{p}$, respectively.\footnote{note
that our definition adequate for groups with involution is
slightly more restrictive than the usual one \cite{M3} } We
consider the set
\eq
\label{b.4}
\wGhe{p}\equiv\Ghe{p}\times\Che{p}
\eqend
with a multiplication $\cdot$ given by
\eq
\label{b.5}
(U,\mu)\cdot (V,\nu)\equiv (UV,\mu\nu^U\chi(U,V))
\eqend
for all $(U,\mu),(V,\nu)\in\wGhe{p}$, with $\chi$ some 2-cochain.
It is easy to see that this multiplication is associative and
allows for an inverse if and only if $\chi$ satisfies
\eq
\label{b.6}
\chi(U,V)\chi(UV,W)=\chi^U(V,W)\chi(U,VW) \nonu
\eqend
for all $U,V,W\in\Ghe{p}$. Moreover, these relations imply that we can
define an involution
\eq
(U,\mu)\mapsto (U,\mu)^*\equiv (U^*,\overline{\mu^{U^*}})
\eqend
making $\wGhe{p}$ to an
unitary group (i.e.\ the inversion is equal to the involution).
Eq.\ \Ref{b.6} are {\em 2-cocycle relations}, and a 2-cochain
$\chi$ satisfying it is a {\em 2-cocycle} of $\Ghe{p}$
\cite{F,FS,M3}.

For every 1-cochain $\beta$ of $\Ghe{p}$, there is an
automorphism $\sig_\beta$ of $\wGhe{p}$ given by
\eq
\label{b.7}
\sig_\beta\left( (U,\mu)\right) \equiv (U,\beta(U)\mu)
\eqend
for all $(U,\mu)\in\wGhe{p}$. It is easy to see that the
$\sig_\beta\left( (U,\mu)\right)$ have product relations similar
to \Ref{b.5} with $\chi$ replaced by $\del\beta\chi$, $\del\beta$
the 2-cochain given by
\eq
\label{b.8}
\del\beta(U,V) \equiv \f{\beta^U(V)\beta(U)}{\beta(UV)}
\eqend
and satisfying the 2-cocycle relation \Ref{b.6}
trivially. A 2-cochain of the form $\del\beta$ \Ref{b.8}, $\beta$
some 1-cochain, is called a {\em 2-coboundary} of $\Ghe{p}$.
Thus it is natural to regard the product \Ref{b.5} equivalent
to all those obtained by replacing $\chi$ with $\del\beta\chi$,
$\del\beta$ any 2-coboundary of $\Ghe{p}$,
and the Abelian extensions $\wGhe{p}$ are in
one-to-one correspondence to equivalence classes of 2-cocycles of
$\Ghe{p}$ which are equal up to a 2-coboundary.

\remark
Obviously the group $\wGhe{p}$ is the {\em semi-direct product} of
$\Che{p}$ by $\Ghe{p}$ with the action of the latter on the former
given by eq.\ \Ref{b.3} \cite{M3}.  There is another Abelian extension
which is the {\em direct product} of $\Che{p}$ with $\Ghe{p}$, and a
corresponding cohomology which is similar (but simpler) than the one
discussed above (see e.g.\ \cite{EL3}). A similar remark applies to
the Lie algebra cohomology discussed below.

Note that the $\mult$-product introduced in Section \ref{sec5} is
naturally associated with this semi-direct product cohomology of
$\Ghe{1}$, whereas the usual operator product is associated with its
direct product cohomology \cite{EL3}.

\subsection{Abelian Extensions of Lie Algebras}
\label{sec3.2}
The Lie algebra of $\Che{p}$ \Ref{b.1} is the set
\eq
\label{b.9}
\che{p}=\map(\Grhe{p};\R)
\eqend
of `smooth' maps $m:\Grhe{p}\to\R;F\mapsto m(F)$ with the Lie bracket
given by
\eq
\ccr{m}{n}(F)= \ccr{m(F)}{n(F)}=0.
\eqend
Corresponding to \Ref{b.3}, the natural representation
$m\mapsto\cL_u m$ of $\ghe{p}$ on $\che{p}$ is given by the Lie
derivative,
\eq
(\cL_u m)(F)\equiv \f{\dd}{\ii\dd t}m(\ee{-\ii tu}F\ee{\ii
tu})\restr{t=0}
\eqend
$\forall m\in\che{p}$, $u\in\ghe{p}$, $F\in\Grhe{p}$.

Similar to the the group case, {\em 1-cochains} and {\em 2-cochains}
of $\ghe{p}$ are `smooth', linear, antisymmetric, $\che{p}$-valued
maps on $\ghe{p}$ and $\ghe{p}\times\ghe{p}$, respectively. The Lie
algebra corresponding to $\wGhe{p}$ is
\eq
\label{b.11}
\wghe{p}\equiv\ghe{p}\oplus \che{p}
\eqend
with the Lie bracket given by
\eqa
\label{b.12}
\CCR{(u,m)}{(v,n)}\equiv \f{\dd}{\ii\dd t}\f{\dd}{\ii\dd s}
(\ee{\ii tu},\ee{\ii tm})\cdot(\ee{\ii sv},\ee{\ii sn})\nonu
\cdot
(\ee{-\ii tu},\ee{-\ii tm})\cdot(\ee{-\ii sv},\ee{-\ii
sn})\restr{s=t=0}.
\eqaend
Using \Ref{b.5}, a simple calculation gives
\eq
\label{b.13}
\CCR{(u,m)}{(v,n)} = (\ccr{u}{v},\cL_u n-\cL_v m + c(u,v) )
\eqend
with
\eq
\label{b.14}
c(u,v)=\f{\dd}{\ii\dd t}\f{\dd}{\ii\dd s}\chi(\ee{\ii tu},\ee{\ii
sv})\chi(\ee{-\ii tu},\ee{-\ii sv})\chi(\ee{\ii tu}\ee{\ii
sv},\ee{-\ii tu}\ee{-\ii sv})\restr{s=t=0},
\eqend
a 2-cochain as \Ref{b.12} implies antisymmetry: $c(u,v)=-c(v,u)$.
It follows from \Ref{b.6} that $c$ obeys the relation
\eqa
\label{b.15}
c(u,\ccr{v}{w})+\cL_u c(v,w) &=& c(\ccr{u}{v},w)-\cL_w c(u,v)
\nonu && + c(v,\ccr{u}{w}) + \cL_v c(u,w)
\eqaend
for all $u,v,w\in\ghe{p}$, which is equivalent to
$\CCR{\cdot}{\cdot}$ fulfilling the Jacobi
identity. Eq.\ \Ref{b.15} is a {\em 2-cocycle relation}, and a
2-cochain satisfying it is a {\em 2-cocycle} of $\ghe{p}$.

The analog of the automorphism $\sig_\beta$ \Ref{b.7} is the map
\eq
\label{b.16}
s_b\left( (u,m)\right) \equiv (u,m+b(u))
\eqend
with $b$ the 1-cochain
\eq
\label{b.16a}
b(u)=\f{\dd}{\ii\dd t}\beta(\ee{i tu})\restr{t=0};
\eqend
this changes $c$ to $c-\dd b$ with
$\dd b$ the 2-cochain given by
\eq
\label{cob}
\dd b(u,v)= b(\ccr{u}{v}) - \cL_u b(v) + \cL_v b(u)
\eqend
and satisfying \Ref{b.15} trivially; such a $\dd b$ is
denoted as {\em 2-coboundary}
of $\ghe{p}$. Similar as in the Lie group case, we regard the Lie
bracket \Ref{b.13} equivalent to the ones obtained by replacing
$c$ with $c-\dd b$, $b$ any 1-cochain.

\section{Quantization of Fermions in External Fields}
\label{sec4}
\subsection{Fermion Field Algebras and Fock Spaces}
\label{sec4.1}
In the spirit of the algebraic approach to quantum field theory
\cite{HK}, we start with the fermion field algebra $\cA(h)$ over
$h$ which contains the observable algebra as subalgebra. $\cA(h)$
is defined as $C^*$-algebra with involution $a\mapsto a^*$,
generated by the elements $\psi^*(f)$, $f\in h$, such that the
mapping $f\mapsto \psi^*(f)$ is linear and the following CAR are
fulfilled
\eqa
\label{2.1}
\car{\psi^{}(f)}{\psi^{*}(g)}&=&\produ{f}{g} \nonu
\car{\psi^{}(f)}{\psi^{}(g)}&=& 0 \qquad \forall f,g\in h
\eqaend
with
\eq
\label{2.2}
\psi^{}(f)=\psi^*(f)^* \quad \forall f\in h
\eqend
($\produ{\cdot}{\cdot}$ denotes the inner product in $h$).

In this paper we consider only unitary representations of $\cA(h)$
on the Fock space $\cF(h)$ over $h$ where the involution $*$ can
be identified with taking the Hilbert space adjoint \cite{BR2} (note
that we use the same symbol $*$ to denote the Hilbert space adjoint
in $h$ and in $\cF(h)$).

Let $a^{(*)}(f)$, $f\in h$ be the annihilation (creation)
operators on $\cF(h)$ satisfying the CAR and
\eq
\label{2.3}
a^{}(f)\Om = 0 \quad \forall f\in h
\eqend
with $\Om$ the vacuum in $\cF(h)$.  We denote as $\cD_{at}(h)$
the set of algebraic tensors in $\cF(f)$ which is the linear span
of the monomials
\[
a^*(f_1)a^*(f_2)\cdots a^*(f_n)\Om
\]
with $f_1,f_2,\cdots f_n\in h$ and $n\in\N_0$. Note that
$\cD_{at}(h)$ is dense in $\cF(h)$.

The free (= Fock-Cook) representation  $\Pi_1$ of $\cA(h)$ on
$\cF(h)$ is given by
\eq
\label{free}
\Pi_1\left( \psi^{(*)}(f) \right)\equiv a^{(*)}(f) \quad \forall f\in h.
\eqend

For all unitary operators $U$ on $h$, there is a unique unitary
operator $\FGam{U}$ on $\cF(h)$ such that
\eq
\label{2.4}
\FGam{U}a^{(*)}(f)\FGam{U}^* = a^{(*)}(Uf) \quad \forall f\in h,
\quad \FGam{U}\Om =\Om ,
\eqend
and one can show that
\eq
\label{2.5}
\FGam{U}\FGam{V}=\FGam{UV}, \quad \FGam{U^*}=\FGam{U}^*
\eqend
for all unitary operators $U,V$ on $h$ \cite{BR2}.  Hence
$\FGam{\cdot}$ provides a unitary representation of the group
$\vec{G}(h)$ of all unitary operators on $h$.  Moreover, for $u$
a self-adjoint operator on $h$, $\FGam{\ee{\ii tu}}$ is a
strongly continuous 1-parameter family of unitary operators on
$\cF(h)$, hence (Stone's theorem \cite{RS2})
\eq
\label{2.6}
\FGam{\ee{\ii tu}} = \ee{\ii t\dFgam{u}}
\eqend
with
\eq
\label{2.7}
\dFgam{u}=\f{\dd}{\ii\dd t}\FGam{\ee{\ii t u}}\restr{t=0}
\eqend
a self-adjoint operator on $\cF(h)$ \cite{BR2}.  The operators
$\dFgam{u}$ are unbounded in general, even if $u$ is bounded.
However, the set
\eq
\label{2.8}
\cD^\infty(h)\equiv
\left\{ \psi\in\cF(h)| \Norm{N^k\psi}<\infty\quad \forall k\in\N
\right\}
\eqend
($\Norm{\cdot}$ the Hilbert space norm) with $N\equiv\dFgam{1}$
the particle number operator, can be shown to be a common, dense,
invariant domain of definition for all $\dFgam{u}$, $u$ a
bounded, self-adjoint operator on $h$. Moreover, it follows from
\Ref{2.5} that
\eq
\label{2.9}
\ccr{\dFgam{u}}{\dFgam{v}}=\dFgam{\ccr{u}{v}}
\eqend
on $\cD^\infty(h)$, i.e.\ that $u\mapsto \dFgam{u}$ provides a
unitary representation of the Lie algebra $\vec{g}(h)$ of
bounded, self-adjoint operators on $h$
\cite{BR2}.

Especially, $\FGam{\cdot}$ and $\dFgam{\cdot}$ by restriction
give representations for all the Lie groups and Lie algebras
$\Map (M^d;G)$ and $\map(M^d;g)$, respectively. However, these
are no highest weight representations, hence they are not very
interesting from a mathematical point of view \cite{KR}.  This
corresponds to the fact that the free representation $\Pi_1$
\Ref{free} of the fermion field algebra $\cA(h)$ is not the one
of interest to quantum field theory: The time evolution is
implemented in this representation by $\FGam{\ee{-\ii tD_A}}$,
and its generator $\dFgam{D_A}$, the multi-particle Hamiltonian,
is not bounded from below \cite{BT4,GL}.

\subsection{Quasi-free Representation}
A physically relevant representation for the model can be
constructed by ``filling up the Dirac sea'' \cite{BT4}: Let
\[
h=h_+\oplus h_-, \quad h_\pm=P^0_\pm h, \quad P^0_\pm =\f{1}{2}(1
\pm \eps), \quad \eps=\sgn{D_A}
\]
be the splitting of the 1-particle Hilbert space in positive and
negative energy subspaces as discussed in Section \ref{sec2.2}.
Then the quasi-free representation $\Pi_\eps$ of $\cA(h)$ is
given by $\Pi_\eps\left(\psi^{(*)}(\cdot)\right) \equiv
\hat{\psi}^{(*)}(\cdot;\eps)$,
\eq
\label{2.10}
\hat{\psi}^{*}(f;\eps)= a^*(P^0_+f)+a^{}(C_\eps P^0_-f)
\quad \forall f\in h
\eqend
with $C_\eps$ a conjugation\footnote{ i.e.\ , $C_\eps$ is
anti-linear and obeys
\protect{\[
\produ{C_\eps f}{C_\eps g}=\produ{g}{f} \quad \forall f,g\in h
\]}
} on $h$ commuting with $\eps$.  Indeed, one can easily check the
multi-particle Hamiltonian in this representation is $\dFgam{|D_A|}$
(with $|D_A|=D_A\eps$) and positive.

Obviously, we can construct a quasi-free representation $\Pi_F$
of $\cA(h)$ on $\cF(h)$ for any grading operator $F$ on $h$.
$\Pi_F$ is called unitarily equivalent to $\Pi_\eps$ if there is
a unitary operator $\cU(F,\eps)$ on $\cF(h)$ such that
\eq
\label{ue}
\cU(F,\eps)\hat{\psi}^*(f;\eps) =
\hat{\psi}^*(f;F)\cU(F,\eps) \quad \forall f\in h .
\eqend
The well-known necessary and sufficient condition for this to
be the case is \cite{KS}
\eq
(F-\eps)\in \I{2},
\eqend
hence $\Grhe{1}$ introduced in the last Section
is just the set of grading operators $F$ with $\Pi_F$ unitarily
equivalent to $\Pi_\eps$.

\remark
The conjugation $C_\eps$ required for the construction of the
quasi-free representation $\Pi_\eps$ \Ref{2.10} is not unique.
Indeed, if $U$ is a unitary operator on $h$ commuting with
$\eps$, then
\[
C_\eps^U\equiv U^{-1}C_\eps U
\]
is a conjugation on $h$ commuting with $\eps$ if (and only if)
$C_\eps$ is.
However, this ambiguity is harmless as the representations obtained
with different choices for
this conjugation are all unitarily equivalent:
Indeed, one easily sees that
\nonueqa
a^*(P^0_+f)+a^{}(C_\eps^U P^0_-f) =
\FGam{P^0_+\oplus C_\eps^UC_\eps P^0_-}
{[}a^*(P^0_+f)+a^{}(C_\eps P^0_-f){]}
\\ \times
\FGam{P^0_+\oplus C_\eps^UC_\eps P^0_-}^* \quad \forall f\in h
\nonueqaend
with $P^0_+\oplus C_\eps^UC_\eps P^0_-$ the unitary operator on
$h$ which is $1$ on $h=P^0_+h$ and $C_\eps^UC_\eps$ on
$h_-=P^0_-h$.

In the following, we find it convenient to fix this ambiguity as
follows: for a given $\eps$ we assume that some conjugation $C_\eps$
is chosen. Then for any unitary operator $U$, the grading operator
$F=U^{-1}\eps U$ has the conjugation $C_F=U^{-1}C_\eps U$.

\subsection{Bogoliubov Transformations}
\label{sec4.3}
Let $F$ be any grading operator on $h$.

Every unitary operator $U$ on $h$ defines an automorphism
$\alpha_U$ of the fermion field algebra $\cA(h)$:
\eq
\alpha_U(\psi^{(*)}(f)) \equiv \psi^{(*)}(Uf)\quad \forall f\in h .
\eqend
Such an $\alpha_U$ is called Bogoliubov transformation.  It is
called unitarily implementable in the quasi-free representation
$\Pi_F$ if there is a unitary operator $\hGam{U;F}$ on $\cF(h)$,
a (so-called) implementer, such that
\eq
\label{ui}
\hGam{U;F}\hat{\psi}^*(f;F)=\hat{\psi}^*(Uf;F)\hGam{U;F}
\quad \forall f\in h .
\eqend
If the implementer $\hGam{U;F}$ exists it is unique up to a phase
\cite{R1}.
The well-known necessary and sufficient criterion for this to be
the case is the Hilbert-Schmidt condition
\cite{R1} (see also the Lemma in Appendix B),
\eq
\ccr{U}{F}\in\I{2},
\eqend
hence $\GhF{1}$ introduced in the last Section can be identified
with the group of all unitarily implementable Bogoliubov
transformations in $\Pi_F$.

\section{Group Structure of the Implementers}
\label{sec5}
\subsection{General Discussion}
Let $F\in\Grhe{1}$. It is known that then the implementers
$\hGam{\cdot;F}$ provide a projective representation of the Lie group
$\Ghe{1}$ with the usual operator product as group multiplication
\cite{EL3}. However, it is possible to define another product,
$\mult$, also providing a group structure of the implementers: From
the defining relations \Ref{2.10} of the quasi-free representation
$\Pi_F$ one can see that every Bogoliubov transformation $\alpha_U$ in
$\Pi_F$ can be written as
\eq
\label{oh}
\hat{\psi}^*(Uf;F)=\FGam{U}\hat{\psi}^*(f;F^U)\FGam{U}^*\quad\forall
f\in h
\eqend
(we just used $UU^*=1$, $U^*FU=F^U$, $U^*C_FU=C_{F^U}$, and eq.\
\Ref{2.4}). Using this formula, eq.\ \Ref{ui}, and assuming
$U,V\in\Ghe{1}$, we rewrite the Bogoliubov
transformation $\alpha_{UV}$ in $\Pi_F$ as follows:
\nonueqa
\hat{\psi}^*(UVf;F)=\FGam{U}\hat{\psi}^*(Vf;F^U)\FGam{U}^* =
\FGam{U}\hGam{V;F^U}\FGam{U}^* \hGam{U;F}\nonu \times
\hat{\psi}^*(f;F)\hGam{U;F}^*\FGam{U}\hGam{V;F^U}^*\FGam{U}^*.
\nonueqaend
{}From this we can see that the unitary operators
$\FGam{U}\hGam{V;F^U}\FGam{U}^*\hGam{U;F}$ and $\hGam{UV;F}$ both
implement the same Bogoliubov transformation $\alpha_{UV}$ in
$\Pi_F$. As the implementer of a Bogoliubov transformation is
unique up to the phase \cite{R1}, we conclude that
\eq
\label{mult}
\hGam{U;F}\mult\hGam{V;F}\equiv
\FGam{U}\hGam{V;F^U}\FGam{U}^*\hGam{U;F}=\nonu
\hGam{UV;F}\chi(U,V;F)
\eqend
with $\chi$ some $\U(1)$-valued function on
$\Ghe{1}\times\Ghe{1}\times\Grhe{1}$.  It is natural to regard
this as {\em definition of a product} $\mult$ of two implementers.
Associativity of this product
\nonueqa
\left(\hGam{U;F}\mult\hGam{V;F}\right)\mult\hGam{W;F} \\=
\FGam{UV}\hGam{W;F^{UV}}\FGam{UV}^*\left(\hGam{U;F}\mult\hGam{V;F}\right)
\\ \stackrel{!}{=}
\hGam{U;F}\mult\left(\hGam{V;F}\mult\hGam{W;F}\right)\\=
\FGam{U}\left(\hGam{V;F^U}\mult\hGam{W;F^U}\right)\FGam{U}^*\hGam{U;F}
\nonueqaend
is equivalent to the following relation
\eq
\label{X.1}
\chi(U,V;F)\chi(UV,W;F)=\chi(V,W;F^U)\chi(U,VW;F)
\eqend
$\forall U,V,W\in\Ghe{1}$ and $F\in\Grhe{1}$.  Similarly, it
follows from \Ref{oh} and \Ref{ui} that for $U\in\Ghe{1}$,
$F\in\Grhe{1}$, the unitary operators
$\FGam{U}^*\hGam{U;F}^*\FGam{U}$ and $\hGam{U^*;F^U}$ both
implement the same Bogoliubov transformation $\alpha_{U^*}$ in
$\Pi_{F^U}$ and therefore are equal up to a phase.  The phases of
the implementers are arbitrary, and we assume them to be chosen
in a `smooth' way (this is possible at least for $U$ in some
neighborhood of the identity --- see below) and such that
\eq
\label{conv}
\hGam{U^*;F^U} = \FGam{U}^*\hGam{U;F}^*\FGam{U},\quad \hGam{1;F} = 1\: .
\eqend
This and \Ref{mult} imply that
\eq
\label{conv1}
\overline{\chi(U,V;F)} =\chi(V^*,V^*;F^{UV}), \quad \chi(U,U^*;F)=1
\eqend
$\forall U,V\in\Ghe{1}$, $F\in\Grhe{1}$.
One can change the phase convention:
\eq
\label{X.2}
\hGam{U;F}\lrar\hGam{U;F}\beta(U;F)
\eqend
with $\beta$ some `smooth',
$\U(1)$-valued function on $\Ghe{1}\times\Grhe{1}$
satisfying
\eq
\label{conv2}
\overline{\beta(U;F)} = \beta(U^*;F^U), \quad \beta(1;F)=1
\eqend
$\forall U\in\Ghe{1}$, $F\in\Grhe{1}$. This amounts to changing
\eq
\label{X.3}
\chi\lrar\del\beta\chi
\eqend
with
\eq
\label{X.4}
\del\beta(U,V;F) \equiv \f{\beta(V;F^U)\beta(U;F)}{\beta(UV;F)}.
\eqend
Moreover, the formula \Ref{CC2} for $\chi$ below shows explicitly that
the mapping $(U,V,F)\mapsto\chi(U,V;F)$ is `smooth' (at least {\em
locally}, i.e.\ for $U,V$ in some neighborhood of the identity) and
satisfies all the relations given above.  Thus the cohomology as
discussed in Section \ref{sec3} naturally emerges here: writing
$\chi(U,V;F)=\chi(U,V)(F)$, one can see that $\chi$ is a 2-cocycle of
the group $\Ghe{1}$ as \Ref{X.1} are just the 2-cocycle relations
\Ref{b.6} and \Ref{conv1} is equivalent to \Ref{222}. Similarly,
$\del\beta$ is a 2-coboundary and \Ref{X.4} and \Ref{conv2} are
equivalent to \Ref{b.8} and \Ref{111}, respectively.
\remark
It is easy to see that the $\mult$- and the operator product are
related to each other: As $\hGam{U;F}\hGam{V;F}$ also implements the
Bogoliubov transformation $\alpha_{UV}$ in $\Pi_F$, there must
be a phase $\eta(U,V;F)$ such that
\[
\hGam{U;F}\mult\hGam{V;F}=\eta(U,V;F)\hGam{U;F}\hGam{V;F} \: .
\]
This allows us to identify
\[
\FGam{U}\hGam{V;F^U}\FGam{U}^*=\eta(U,V;F)\hGam{U;F}\hGam{V;F}\hGam{U;F}^*
\]
for all $U,V\in\Ghe{1}$, $F\in\Grhe{1}$. One can regard $\eta$ as an
intertwiner relating the semidirect-product and the direct-product
cohomologies of $\Ghe{1}$ (see the remark in Section \ref{sec3.1}).
Under a change of phases of the implementers \Ref{X.2} it obviously
transforms as
\[
\eta(U,V;F)\lrar\f{\beta(V;F)}{\beta(V;F^U)}\eta(U,V;F).
\]
It is even possible to choose the phases of the implementers
such that
\[
\eta(U,V;F)=1,
\]
i.e.\ that the $\mult$- and the operator product of the
implementers coincide (below we show this explicitly for
$U,V\in\Ghe{1}$ in some neighborhood of the identity).

Hence at this stage, the two products are essentially the same.
However, when introducing the implementers $\hpGam{U;F}$ for
$U\in\Ghe{p}$, $F\in\Grhe{p}$, $p>1$, this will be no longer the case:
it is only the $\mult$-product that allows for a multiplicative
regularization leading to an {\em associative} product of these
implementers, whereas all non-trivial regularizations of the operator
product are non-associative and give rise to a non-trivial 3-cocycle
\cite{C}. Indeed, a non-trivial multiplicative regularization of the
operator product would provide a ``highest weight'' representation of
a non-trivial {\em central} extension of $\Ghe{p}$ which does not
exist for $p>1$ \cite{P}.

\subsection{Explicit Formulas}
\label{sec5.2}
The explicit form of the implementers $\hGam{U;F}$ for
$U\in\GhF{1}$, $F$ any grading operator on $h$, was worked out by
Ruijsenaars \cite{R1}, and is quite complicated in general.
Assuming, however, that $P_-UP_-$ has a bounded inverse on
$P_-h$, it simplifies and can be used to explicitly evaluate the
2-cocycle $\chi$ in \Ref{mult}.

We denote the set of all unitary operators $U$ with $P_{-}UP_{-}$
bijective on $P_-h$ as $\UhF$, and $\UheF=\Uhe\cap\UhF$. Note that
for all $F\in\Grhe{p}$, $\Ghe{p}\cap\UheF$ is a neighborhood (but no
subgroup) of $\Ghe{p}$ for all $p\in\N_0$.

In Appendix A we prove that for all $F\in\Grhe{1}$,
$U,V\in\Ghe{1}\cap\UheF$, the phases of the implementers can be chosen
such that
\eq
\label{CC2}
\chi(U,V;F) = \left(
\f{\det{1+[P_-^0UP_-^0]^{-1}P_-^0UP_+^0VP_-^0[P_-^0VP_-^0]^{-1} }}%
{\det{1+[P_-^0V^*P_-^0]^{-1}P_-^0V^*P_+^0U^*P_-^0[P_-^0U^*P_-^0]^{-1}
}}
\right)^{1/2}
\eqend
is independent of $F$. Moreover, we show that if in addition
$U,V\in\Ghe{0}$, one can write this locally as 2-coboundary:
$\chi=\del\beta_1$, with
\eq
\label{CB2}
\beta_1(U;F)=\left(\f{\det{P^0_+ + {[}P^0_-U P^0_-{]}^{-1} }}%
{\det{P^0_+ + {[}P^0_-U^* P^0_-{]}^{-1} }}\right)^{1/2}
\eqend
demonstrating that $\chi$ is locally a trivial 2-cocycle for $\Ghe{0}$
but (as $\beta_1(U;F)$ does not exist for general
$U\in\Ghe{1}\cap\UheF$) non-trivial for $\Ghe{1}$ .

\subsection{Interpretation}
\label{sec5.3}
For each $F\in\Grhe{1}$ we have a quasi-free representation $\Pi_F$ of
the fermion field algebra $\cA(h)$ on the Fock space $\cF(h)$, and
this can also be regarded as a representation of $\cA(h)$ on a Fock
space bundle with the base space $\Grhe{1}$ and fibres $\cF(h)$.  Eq.\
\Ref{oh} shows that it is natural to interpret $\hat{\Gam}(U;F)^*$ as
transformation from $\Pi_{F^U}$ to $\Pi_U$. This suggests to
define a mapping $\hGam{U}$ on the space $\She{1}$ of sections in the
above-mentioned bundle.  To be specific, $\She{1}$ is the vector space
of `smooth' mappings
\[
\Psi:\Grhe{1}\to\cF(h),F\mapsto \Psi(F),
\]
and it is a $\CChe{1}$-module \cite{H} with
\eq
\label{x.3}
\CChe{1}\equiv\Map(\Grhe{1};\C)
\eqend
the ring of `smooth', $\C$-valued functions on $\Grhe{1}$:
\eq
\label{x.4}
(\nu\Psi)(F)=(\Psi\nu)(F)\equiv \nu(F)\Psi(F)\quad
\forall\nu\in\CChe{1},\Psi\in\She{1}.
\eqend
Moreover, one can define an ``inner product''\footnote{to be precise, a
$\CChe{1}$-valued Hermitian sesquilinearform},
\eq
\label{x.5}
<\!<\Psi_1,\Psi_2>\!>(F)\equiv\Produ{\Psi_1(F)}{\Psi_2(F)}\quad
\forall\Psi_1,\Psi_2\in\She{1}.
\eqend
Then if $\hGam{U}$ is defined as\footnote{I am grateful to S. N. M.
Ruijsenaars for suggesting this to me}
\eq
\label{x.6}
(\hGam{U}\Psi)(F)\equiv \hat{\Gamma}(U;F)^*\FGam{U}\Psi(F^U)
\quad\forall\Psi\in\She{1} \: ,
\eqend
the definition \Ref{mult} of the $\mult$-product implies
\eq
\label{x.8}
(\hGam{U}\mult\hGam{V})\Psi \equiv \hGam{U}(\hGam{V}\Psi)=
\overline{\chi(U,V)}\, \hGam{UV}\Psi
\eqend
$\forall\Psi\in\She{1},U,V\in\Ghe{1}$. Moreover, it follows from
\Ref{conv} the involution $*$ is identical with the adjungation
\eq
<\!<\Psi_1,\hGam{U}\Psi_2>\!>(F)=<\!<\hGam{U}^*\Psi_1,\Psi_2>\!>(F^U)\quad
\forall\Psi_1,\Psi_2\in\She{1} \: ,
\eqend
and the $\hGam{U}$ are unitary with respect to $<\!<\cdot,\cdot>\!>$.
Thus $(U,\nu)\mapsto \hGam{U}\nu$ provides a ``unitary''
representation of an Abelian extension $\widehat{\vec{G}_1}(h;\eps)$
of $\Ghe{1}$ on the space $\She{1}$ of sections in a Fock-space
bundle, and the corresponding 2-cocycle is {\em determined} by
\Ref{mult} (and equal to $\overline{\chi(U,V;F)}$ \Ref{CC2} in
some neighborhood of the identity).

\remark
As we can choose the 2-cocycle $\chi$ in \Ref{mult} $F$-independent,
$\widehat{\vec{G}_1}(h;\eps)$ is equivalent a central extension of
$\Ghe{1}$. The latter is the same playing a prominent role in the
theory of loop groups as mentioned in the introduction (this can be
explicitly seen by combining our results here with the ones from Ref.\
\cite{EL3}).

\section{Current Algebras in $(1+1)$-Dimensions}
\label{sec6}
\subsection{General Discussion}
\label{sec6.1}
{}From the results of the last Section one can easily obtain the
corresponding formulas on the Lie algebra level.

For $F\in\Grhe{1}$ and $u\in\ghe{1}$, the current $\dhgam{u;F}$
can be defined as
\eq
\label{current}
\dhgam{u;F}=\f{\dd}{\ii\dd t}\hGam{\ee{\ii tu};F}\restr{t=0}.
\eqend
Indeed, it was shown by Carey and Ruijsenaars \cite{BT4} that for
any grading operator $F$ on $h$ and all $u\in\ghF{1}$, there is a
self-adjoint operator $\dhgam{u;F}$ on $\cF(h)$ such that
$\ee{\ii t\dhgam{u;F}}$ for all $t\in\R$ implements the
Bogoliubov transformation $\alpha_{\ee{\ii tu}}$ in $\Pi_F$, and
it therefore must be equal to $\hGam{\ee{\ii tu};F}$ up to a
`smooth' function $\tilde{\eta}_F:\R\to \U(1); t\mapsto
\tilde{\eta}_F(\ee{\ii tu})$.  Moreover, $\cD^\infty(h)$
\Ref{2.8} is a common, dense, invariant domain for all
$\dhgam{u;F}$, $u\in\ghe{1}$, hence by Stone's theorem
\cite{RS2}, the differentiation in \Ref{current} is defined in
the strong sense on $\cD^\infty(h)$.  We define the Lie bracket
of the currents as\footnote{note that this is well-defined on
$\cD^\infty(h)$}
\eqa
\CCR{\dhgam{u;F}}{\dhgam{v;F}}\equiv \f{\dd}{\ii\dd t}
\f{\dd}{\ii\dd s}\hGam{\ee{itu};F}\mult\hGam{\ee{isv};F}\nonu
\mult\hGam{\ee{-itu};F}\mult\hGam{\ee{-isv};F}\restr{s=t=0},
\eqaend
and with \Ref{mult} we obtain
\eq
\label{ca1}
\CCR{\dhgam{u;F}}{\dhgam{v;F}}=\dhgam{\ccr{u}{v};F}+c(u,v;F)
\eqend
with the Schwinger term $c$ a 2-cochain of the Lie algebra
$\ghe{1}$ given by eq.\ \Ref{b.14} with $\chi$
\Ref{CC2}, and therefore satisfying the 2-cocycle relation \Ref{b.15}.
In Appendix C we show that
\eq
\label{KP}
c(u,v;F)= \f{1}{4}\tra{\ccr{u}{\eps}\ccr{v}{\eps}\eps} \: .
\eqend
This is the cocycle derived originally by Lundberg \cite{Lund}
(in a different but equivalent form)
and which is usually referred to as Kac-Peterson cocycle \cite{PS}.

Changing the phases of the implementers \Ref{X.2}
changes
\eq\label{X.A}
\dhgam{u;F}\lrar\dhgam{u;F}+b(u;F)
\eqend
with $b$ the 1-cochain of $\ghe{1}$ given by
\Ref{b.16a}, and by an explicit calculation one can check that this amounts
to changing
\eq
\label{X.B}
c\lrar c-\dd b
\eqend
with $\dd b$ the 2-coboundary \Ref{cob}.
Thus $\CCR{\cdot}{\cdot}$ is not just the commutator. It is
natural to set
\eq
\label{X.C}
\CCR{\dhgam{u;F}}{n(F)}\equiv (\cL_u n)(F)
\eqend
for all `smooth', $\C$-valued functions $n$ on $\Grhe{1}$ as this
definition naturally accounts for \Ref{X.A}, \Ref{X.B} and the Jacobi
identity in \Ref{b.15}.

\remark
As discussed above, in the phase convention leading to the
$F$-independent 2-cocycle \Ref{CC2}, the $\mult$-product
coincides with the operator product; therefore, the Lie bracket of
the currents in \Ref{ca1} is equivalent to (but not identical
with) the commutator of these as operators on $\cF(h)$.

\subsection{Interpretation}
\label{sec6.2}
Corresponding to the interpretation of $\hGam{U;F}$ as mapping
$\hGam{U}$ on the space $\She{1}$ of sections in a Fock space
bundle in Section \ref{sec5.3}, it is natural to set
\eq
\label{x.9}
\dhgam{u}\Psi\equiv\f{\dd}{\ii\dd t}\hGam{\ee{\ii tu}}\Psi\restr{t=0}
\quad \forall \Psi\in\She{1}
\eqend
for all $u\in\ghe{1}$. This defines the currents $\dhgam{u}$ as
mappings on $\She{1}$. Corresponding to \Ref{x.8}, the Lie bracket of
two such currents is
\eq
\label{x.10}
\CCR{\dhgam{u}}{\dhgam{v}}\Psi = \ccr{\dhgam{u}}{\dhgam{v}}\Psi
\quad\forall\Psi\in\She{1}
\eqend
with $\ccr{\cdot}{\cdot}$ the usual commutator. With \Ref{x.8} we
obtain\footnote{the minus sign arises due to the complex conjugation}
\eq
\label{x.11}
\CCR{\dhgam{u}}{\dhgam{v}}=\dhgam{\ccr{u}{v}}-c(u,v)
\eqend
$\forall u,v\in\ghe{1}$, showing that $(u,n)\mapsto\dhgam{u}+n$
provides a ``unitary'' representation of the Abelian extension
$\widehat{\vec{g}_1}(h;\eps)$ associated with the 2-cocycle $c$
\Ref{KP} on $\She{1}$.

Eq.\ \Ref{x.9} and \Ref{x.10} give a simple interpretation to the Lie
bracket $\CCR{\cdot}{\cdot}$ in eq.\ \Ref{ca1} (and its seemingly
strange property \Ref{X.C}): From \Ref{x.9} and \Ref{x.6}  we obtain
\[
(\dhgam{u}\Psi)(F)=
\f{\dd}{\ii\dd t}
\hGam{\ee{\ii tu};F}^*\FGam{\ee{\ii tu}}
\Psi(\ee{-\ii tu}F\ee{\ii tu})\restr{t=0}\: ,
\]
hence
\eq
\label{x.12}
(\dhgam{u}\Psi)(F) = \left(\cL_u + \tilde{J}(u;F)\right)\Psi(F)
\eqend
with
\eq
\label{x.13}
\tilde{J}(u;F)\equiv \f{\dd}{\ii\dd t} \tilde{\Gam}(\ee{\ii tu};F)^*
\FGam{\ee{\ii tu}} \restr{t=0}\: = \dFgam{u}-\dhgam{u;F}\: .
\eqend
This shows that it is natural to regard $\dhgam{u}$ as Gauss' law
generators rather than as currents, and to interpret \Ref{x.11} as an
abstract version of the algebra of the Gauss' law constraint operators
in $(1+1)$-dimensions
\cite{Jackiw}.

\section{Multiplicative Regularization}
\label{sec7}
\subsection{General Discussion}
\label{sec7.1}
For $U\in\Ghe{1}$, the implementer $\hGam{U;F}$ can be written as
\cite{R1}
\eq
\label{imp}
\hGam{U;F}=E(U;F)N(U;F)
\eqend
with $E(U;F)$ an operator on $\cF(h)$ evaluated such that it
implements the Bogoliubov transformation $\alpha_U$ in $\Pi_F$
(i.e.\ obeys
\Ref{ui}) and $N(U;F)\in\Cb$ a
normalization constant needed to make the implementer unitary.
It is given (up to a phase) by the condition that
$\hGam{U;F}\hGam{U;F}=1$, i.e.\
\[
\norm{N(U;F)}^{-2} =\Produ{\Om}{E(U;F)E(U^*;F)\Om}.
\]
By the explicit formula for $E(U;F)$ given in eq.\ \Ref{A.2} for
$U\in\Ghe{1}\cap\UheF$ one finds
\eqa
\label{N}
N(U;F)=\left(\det{1-P_-UP_+U^*P_-}
\f{\detz{1+a(U;F)}}{\detz{1+a(U;F)}^*}
\right. \nonu  \times\left.
\f{\exp{(\tra{P_+^0a(U;F)P_+^0 +
P_-^0a(U;F)P_-^0})}}{\exp{(\tra{P_+^0a(U;F)P_+^0 +
P_-^0a(U;F)P_-^0}^*)}} \right)^{1/2}
\eqaend
with $a(U;F)$ defined in Appendix A, eq.\ \Ref{defa} (we also show
there that \Ref{N} is well-defined).

The crucial point is that $E(U;F)$ does not
only exist for $U\in\Ghe{1}$, but that it can be defined in as
sesquilinear form \cite{Kato} on the set $\cD_{at}(h)$ of
algebraic tensors in $\cF(h)$ (cf.\ Section
\ref{sec4.1}) for
all $U\in\GhF{\infty}$. Hence the non-existence of $\hGam{U;F}$ for
general $U\in\Ghe{p}$, $p>1$, is only due to the non-existence of the
normalization constant $N(U;F)$ (this observation is due to
Ruijsenaars \cite{R2,R4}).  As $E(U;F)$ for $U\in\GhF{\infty}$ obeys
\Ref{ui} in the form sense on $\cD_{at}(h)$, it is natural to perform
a multiplicative regularization of the implementers
\eq
\label{reg1}
\hGam{U;F}\lrar \hpGam{U;F}
\eqend
and to define\footnote{this and similar eqs.\ below have to be
interpreted as follows: for each sequence in $\Ghe{1}$ converging to
$U\in\Ghe{p}$ in the $\NORM{\cdot}_p$-norm, the r.h.s. has a
well-defined limit depending only on $U$, and the l.h.s. is defined as
this limit}
\eq
\label{hGamp}
\hpGam{U;F}\equiv \hGam{U;F}\beta^N_p(U;F)
\eqend
with $\beta^N_p(U;F)$ such that it cancels the divergency in $N(U;F)$
for all $U\in\Ghe{p}$, $F\in\Grhe{p}$. This regularization, however,  is
not sufficient due to the fact that the operator product of forms
cannot be defined in general. Explicitly we find for
$F\in\Grhe{1}$, $U,V\in\Ghe{1}\cap\UheF$,
\eq
\label{N.4}
E(U;F)\mult E(V;F)=\cE(U,V;F)E(UV;F),
\eqend
with (see Appendix A, \ eq.\ \Ref{A.9a})
\eq
\label{N.5}
\cE(U,V;F)=\det{1+[P_-UVU^*P_-]^{-1}P_-UVU^*P_+UP_-[P_-UP_-]^{-1}}\: .
\eqend
This does not exist (diverges) for $U,V\notin\Ghe{1}$.\footnote{ one
obtains a similar formula for the operator product --- see
\cite{EL3}} Therefore, $\mult$ is not the appropriate group
multiplication for the implementers $\hpGam{U;F}$, and we are forced
to regularize it as well:
\eq
\label{reg}
\mult\lrar\multp .
\eqend
{}From a quantum field theoretic point of view, it is natural to demand
that the regularizations \Ref{hGamp} and \Ref{reg} are {\em minimal}
in a sense made precise below.

To be specific: Using the Lemma in Appendix B, it is easy to see that
{\em locally} (i.e.\ for $U,V\in\Ghe{p}\cap\UheF$ and $F\in\Grhe{p}$),
$P_-UP_+U^*P_- \in\I{p}$, $a(U;F)\in\I{2p}$ (cf. \Ref{defa}), and the
r.h.s. of eq.\ \Ref{N.5} is formally of the form $\det{1+(\cdots)}$
with $(\cdots)\in\I{p}$. As discussed in Section \ref{sec2.3}, a
minimal regularization of the implementers \Ref{reg1} therefore
amounts to dropping the non-existent traces in \Ref{N}, especially
replacing the determinants in by regularized ones as follows:
\eqa
N(U;F)\lrar N_p(U;F)\hspace*{4cm} \nonu \equiv
\left( \detp{1-P_-UP_+U^*P_-}
\f{\detzp{1+a(U;F)}}{\detzp{1+a(U;F)}^*} \right)^{1/2} \: ,
\eqaend
and to define $\hpGam{U;F}$ \Ref{hGamp} with
\eq
\label{betap}
\beta^N_p(U;F)=\f{N_p(U;F)}{N(U;F)}\: .
\eqend
However, in order to regularize the $\mult$-product, we cannot simply
replace in \Ref{N.5} the $\det{\cdot}$ by $\detp{\cdot}$ as {\em the
regularized product should be associative}. Thus the (non-trivial)
problem is to find an appropriate function
$\beta_p^\mult:\Ghe{1}\times\Grhe{1}\to\Cb$ diverging for general
$U,V\in\Ghe{p}$ and $F\in\Grhe{p}$ such that
\eq
\label{betam}
\hpGam{U;F}\multp\hpGam{V;F}\equiv
\hpGam{U;F}\mult\hpGam{V;F}
\f{\beta^\mult_p(V;F^U)\beta^\mult_p(U;F)}{\beta^\mult_p(UV;F)}
\eqend
exists and is equal to $\hpGam{UV;F}$ up to a phase, as such (and
only such) a regularization maintains associativity and leads to
a projective representation of the group $\Ghe{p}$. Then
\eq
\label{multp}
\hpGam{U;F}\multp\hpGam{V;F}=\hpGam{UV;F}\chi_p(U,V;F)
\eqend
with
\eq
\label{chip}
\chi_p=\del\beta_p \chi,\quad \beta_p\equiv \beta^N_p\beta^\mult_p
\eqend
and $\del$ defined in \Ref{X.4}.

Our discussion above can be summarized by saying that we have to find
a 1-cochain $\beta_p$ of the group $\Ghe{1}$ such that $\chi_p$
\Ref{chip} extends to a (non-trivial) 2-cocycle of the group
$\Ghe{p}$. Moreover (and this can be regarded as definition of {\em
minimal regularization}), $\beta_p$ should be such that locally
\eqa
\label{regu}
\chi_p(U,V;F)=
r_p(U,V;F)\hspace*{4cm} \nonu\times \left(
\f{\detp{1+[P_-^0UP_-^0]^{-1}P_-^0UP_+^0VP_-^0[P_-^0VP_-^0]^{-1} }}%
{\detp{1+[P_-^0V^*P_-^0]^{-1}P_-^0V^*P_+^0U^*P_-^0[P_-^0U^*P_-^0]^{-1}
}}
\right)^{1/2}
\eqaend
with $r_p$ some 2-cochain of $\Ghe{p}$.  Obviously, if such a
1-cochain $\beta_p$ exists, it is locally unique up to a 1-cochain of
$\Ghe{p}$, hence the minimal regularization and the 2-cocycle
$\chi_p$ of $\Ghe{p}$ in \Ref{chip} are locally unique up to a
2-coboundary.

\remark
It is easy to see that the $\hpGam{U;F}$ for $p>1$ are in general
sesquilinear forms only and cannot be promoted to (unbounded)
operators on $\cF(h)$.  Indeed, for $U\in\GhF{1}$ it follows from
\Ref{hGamp} and the unitarity of $\hGam{U;F}$ that
\[
\Produ{\hpGam{U;F}\psi}{\hpGam{U;F}\psi}=\norm{\beta_p^N(U;F)}^2
\Produ{\psi}{\psi} \quad \forall\psi\in\cF(h),
\]
and this does not exist for general $U\in\GhF{p}$.

\subsection{Explicit Result for $p=2$}
\label{sec7.2}
For $p=2$,
the 1-cochain providing the minimal regularization as
discussed above is given by
\eq
\label{CB4}
\beta_2(U;F)=
\exp{(\f{1}{4}\tra{P_-^0 F P_+^0 U P_-^0 U^* P_-^0
 -P_-^0 U P_-^0 (U^*) P_+^0 F P_-^0  })}
\eqend
(by using the Lemma in Appendix B, it is easy to see that this is
indeed a 1-cochain of $\Ghe{1}$).
The proof of this result is given in Appendix D.

The results of \cite{MR,FT} show that the 1-cochains $\beta_p$ of
$\Ghe{1}$ with the properties discussed in the last Section exist for
all $p>1$.

\subsection{Interpretation}
Similar as discussed in Section \ref{sec5.3} for the case $p=1$, it is
natural to regard the implementers $\hpGam{U;F}^*$, $U\in\Ghe{p}$,
$F\in\Grhe{p}$, as mapping $\hpGam{U}$ on the set $\She{p}$ of
sections in the Fock space bundle with the base space $\Grhe{p}$ and
the fibres $\cF(h)$ carrying a module structure and defined completely
analogous to $\She{1}$. The $\multp$-product can then be understood by
a multiplicative regularization of \Ref{x.8}. Then
\eq
\hpGam{U}\multp\hpGam{V}=\overline{\chi_p(U,V)}\, \hpGam{UV}\: ,
\eqend
and $(U,\nu)\mapsto\hpGam{U}\nu$ provides a representation of the
Abelian extension $\widehat{\vec{G}_p}(h;\eps)$ of $\Ghe{p}$
associated with the 2-cocycle $\overline{\chi_p}$ on $\She{p}$. In
contrast to the case $p=1$, the $\hpGam{U}$ for $p>1$ are intertwiners
between unitary inequivalent representations of the fermion field
algebra $\cA(h)$.

\section{Current Algebras in Higher Dimensions}
\label{sec8}
\subsection{Formal Construction}
On the Lie algebra level, the currents $\dhpgam{u;F}$ for
$F\in\Grhe{p}$ and $u\in\ghe{p}$ can be formally defined as
\eq
\label{X.10}
\dhpgam{u;F}\equiv \f{\dd}{\ii\dd t}\hpGam{\ee{itu};F}\restr{t=0},
\eqend
and the (regularized) Lie bracket for such currents is
\eqa
\label{X.11}
\CCR{\dhpgam{u;F}}{\dhpgam{v;F}}_p\equiv \f{\dd}{\ii\dd t}
\f{\dd}{\ii\dd s}\hpGam{\ee{itu};F}\multp\hpGam{\ee{isv};F}\nonu
\multp\hpGam{\ee{-itu};F}\multp\hpGam{\ee{-isv};F}\restr{s=t=0}\: .
\eqaend
Using \Ref{multp}, this results in (cf.\ Section \ref{sec3.2})
\eq
\label{X.12}
\CCR{\dhpgam{u;F}}{\dhpgam{v;F}}_p=\dhpgam{\ccr{u}{v};F}+c_p(u,v;F)
\eqend
with $c_p=c$ given by eq.\ \Ref{b.14} with $\chi=\chi_p$
\Ref{chip}. It follows that
\eq
\label{X.13}
c_p=c-\dd b_p
\eqend
with $b_p=b$ given by eq.\ \Ref{b.16a} for $\beta=\beta_p$.
Eq.\ \Ref{X.12} provides the abstract current algebra
in $(d+1)$-dimensions, $d=2p-1$.

\subsection{Explicit Results for $p=2$}
For $p=2$ we obtain from \Ref{CB4}
\eq
\label{cob4}
b_2(u;F) = -\f{1}{16}\tra{\ccr{u}{\eps}\ccr{F}{\eps}},
\eqend
and $c_2=c-\dd b_2$ results in
\eq
\label{MR}
c_2(u,v;F)= \f{1}{8}\tra{\ccr{\ccr{u}{\eps}}{\ccr{v}{\eps}}(\eps -F)}
\eqend
(the proof is contained in Appendix E).
This is exactly the cocycle of  Mickelsson and Rajeev \cite{MR}.

\subsection{Existence of the Currents}
The discussion above is formal because we did not give
a precise mathematical meaning to the differentiation
in \Ref{X.10}. In this subsection we complete our argument in this
respect by showing that $\dhpgam{u;F}$ is defined as
sesquilinear form on $\cD_{at}(h)$
for all $u\in\ghe{p}$, $F\in\Grhe{p}$, $p>1$.

It follows from the result of Carey and Ruijsenaars \cite{BT4} that
\eq
\label{X.14}
\dd E(u;F) = \f{\dd}{\ii\dd t} E(\ee{\ii tu};F)\restr{t=0}
\eqend
($E$ is given in Appendix A, eq.\ \Ref{A.2})
exists in the strong sense on $\cD^\infty(h)$
\Ref{2.8} for all $u\in\ghF{1}$, $F$ any grading operator on $h$,
and it is equal to
\eq
\label{X.15}
\dd E(u;F) = \dFgam{P_+uP_+}-\dFgam{P_-u^tP_-}+u_{+-}a^*a^* +u_{-+}aa
\eqend
($u^t=C_Fu^*C_F$) with $u_{\pm\mp}=P_{\pm}uP_{\mp}$,
$\dFgam{\cdot}$ \Ref{2.7}, and $u_{+-}a^*a^*$, $u_{-+}aa$ defined
in Appendix A, eqs.\ \Ref{Aaa}{\em ff}. Hence by \Ref{imp}
and \Ref{current}
\eq
\label{X.16}
\dhgam{u;F}=\dd E(u;F)+n(u;F)
\eqend
with $n=b$ given by \Ref{b.16a} for $\beta=N$ \Ref{N}. By explicit
calculation
\eqa
\label{X.17}
n(u;F)=\tra{u(P_- -P_-^0)} =
\nonu
\f{1}{2}\tra{P_+^0u(\eps-F)P_+^0+ P_-^0u(\eps-F)P_-^0}
\eqaend
(note that $\tra{a}=\tra{P_+^0aP_+^0 +P_-^0aP_-^0}$ for $a\in\I{1}$).
For $F\in\Ghe{1}$, $u\in\Grhe{1}$, the r.h.s. of this is obviously
finite (cf.\ the Lemma in Appendix B).

Now it is easy to see that thought $u_{+-}a^*a^*$ and $u_{-+}aa$
exist as operators on $\cF(h)$ only if $u_{\pm\mp}\in\I{2}$ (cf.\
\cite{BT4}), they exist as sesquilinear forms on $\cD_{at}(h)$
whenever $u_{\pm\mp}$ are compact operators. Hence $\dd E(u;F)$
\Ref{X.15} exists as form on $\cD_{at}(h)$ for all
$u\in\ghF{\infty}$, and we {\em define} the r.h.s. of \Ref{X.14}
to be equal to the r.h.s.  of \Ref{X.15} for all
$u\in\ghF{\infty}$. Then it follows from \Ref{X.10},
\Ref{imp} that
\eq
\label{X.18}
\dhpgam{u;F} = \dhgam{u;F} + b_p^N(u;F)
\eqend
with $b_p^N=b$ given by \Ref{b.16a} for $\beta=\beta^N_p$ \Ref{betap},
and this exists as form on $\cD_{at}(h)$ as $(n+b^N_p)(u;F)$ is (by
construction) finite for all $u\in\ghe{p}$, $F\in\Grhe{p}$.
Moreover, \Ref{X.11} is equivalent to
\eq
\label{X.19}
\CCR{\dhpgam{u;F}}{\dhpgam{v;F}}_p =
\CCR{\dhpgam{u;F}}{\dhpgam{v;F}} - \dd b^\mult_p(u,v;F)
\eqend
with $b_p^\mult=b$ \Ref{b.16a} for $\beta=\beta^\mult_p$ \ (cf.\
\Ref{betam}) (of course, this again gives \Ref{X.12} and
\Ref{X.13} with $b_p=b^N_p+b^\mult_p$).

\subsection{Complexification}
The complexification $\vec{g}_p^c(h;\eps)$ of $\ghe{p}$ is
the Lie algebra of {\em all} bounded operators $u$ on $h$
obeying $\ccr{u}{F}\in\I{p}$. Obviously,
\eq
\dhpgam{u;F}\equiv \dhpgam{\f{u+u^*}{2};F}+\ii \dhpgam{\f{u-u^*}{2\ii};F}
\eqend
is well-defined as form on $\cD_{at}(h)$, and it follows that
$u\mapsto\dhpgam{u;F}$ is linear and the relations \Ref{X.12} are
fulfilled for all $u,v\in\vec{g}_p^c(h;\eps)$. Hence we have in
fact a representation of the complexification
$\widehat{\vec{g}_p^c}(h;\eps)$ of $\widehat{\vec{g}_p}(h;\eps)$.

\subsection{Interpretation}
The interpretation of the abstract current algebra in Section
\ref{sec6.2} generalizes trivially from $(1+1)$ to $(d+1)$-dimensions,
$d=2p-1\geq 3$, and we have a representation of the Abelian extension
$\widehat{\vec{g}_p}(h;\eps)$ of $\ghe{p}$ associated with the
2-cocycle $c_p$ on the space $\She{p}$.  Introducing
\eq
\cG_p(u;F)\equiv \cL_u + \tilde{J}_p(u;F)
\eqend
with $\tilde{J}_p$ defined similarly as $\tilde{J}$ in eq.\
\Ref{x.13}, we can write \Ref{X.12} as
\eq
\ccr{\cG_p(u;F)}{\cG_p(v;F)}_p = \cG_p(\ccr{u}{v};F) - c_p(u,v;F)
\eqend
where $\ccr{\cdot}{\cdot}_p$ is the regularized commutator defined
similarly as $\CCR{\cdot}{\cdot}_p$ in \Ref{X.19}. We suggest to
regard this as general, abstract version of the anomalous commutator
relations of the Gauss' law constraint operators in
$(d+1)$-dimensions, $d=2p-1$, as discussed by Faddeev and Shatiashvili
\cite{F,FS}.
\section{Final Comments}
\label{sec9}

One can expect that representations of the group $\vec{G}_2$ and
its Lie algebra $\vec{g}_2$ as studied in this paper could give a
non-perturbative understanding of all ultra-violet divergencies
arising in the fermion sector of Yang-Mills theory with fermions in
$(3+1)$-dimensions. Indeed, not only the implementers of the gauge
transformations, but many other (probably all) fermion operators
of interest in these theories can be regarded as a second
quantization of operators in $\vec{G}_2$ or $\vec{g}_2$, no matter
whether one has Weyl- or massive or massless Dirac fermions.

For example, it is known that the Dirac operator $D_{A(t)}$ for an
arbitrary\footnote{reasonably smooth} time-dependent external
Yang-Mills field $A(t)$ is in $\vec{g}_2$ (for $\eps=\sgn{D_A}$, $A$
an arbitrary Yang-Mills field configuration), and the time-evolution
operators $u(s,t)=T\exp{(-i\int_s^t\dd{r}A(r))}$ generated by
$D_{A(t)}$ all are in $\vec{G}_2$ (see \cite{M5} and references
therein). Hence, using the results of this paper, one can give an
explict construction of the time evolution operators $\cU(s,t)$ for
fermions interacting with $A(t)$, as family of sesqulinear forms which
are a 1-paramter group with repect to the $\multz$-product, i.e.\ obey
the relations
\eq
\label{l1}
\cU(t,t)=1, \quad
\cU(r,s)\multz \cU(s,t) = \cU(r,t) \quad \forall r,s,t\in\R\: .
\eqend
Indeed, it is natural to set (cf. Section \ref{sec5.3})
\eq
\cU(s,t) \equiv \gamma(s,t)\hzGam{u(s,t);F(s)}
\eqend
with $F(s)=u(s,t_0)\eps u(t_0,s)$, $t_0$ is arbitrary,\footnote{for
$A(t)\to 0$ sufficiently fast for $t\to-\infty$, a natural choice
would be $t_0=-\infty$ and $\eps=\sgn{D_0}$} and determine the phases
$\gamma(s,t)$ such that the relations \Ref{l1} are fulfilled. From
\Ref{multp} for $p=2$ one then gets a condition with an
essentially unique solution and which can be solved by a
technique similar to one described in \cite{EL3}.  For Dirac fermions
one has, of course, $\gamma(s,t)=1$ $\forall s,r\in\R$, but for
Weyl fermions one gets a non-trivial solution.  The
latter can also be used for a simple, direct construction of the
non-trivial phase of the $S$-operator of Weyl fermions in external,
time-dependent Yang-Mills fields alternative to the one given recently
by Mickelsson \cite{M5} (we are planning to report on that in more
detail in a future publication).

We finally remark that corresponding results for the Lie group
$\vec{G}_1$ and Lie algebra $\vec{g}_1$ are already sufficient for a
complete understanding of $(1+1)$-dimensional gauge theories with
fermions.  However, this crucially relies on the fact that Yang-Mills
fields on a cylinder have only a finite number of physical degrees of
freedom with an (essentially) unique Hilbert space representation and
therefore do not lead to divergencies (see e.g.\
\cite{M4}). Moreover, it is possible to eliminate all
gauge degrees of freedom and to explictly construct all physical
states \cite{LS1,LS2}.  In higher dimension, a Yang-Mills field has an
infinite number of physical degrees of freedom and associated
divergencies which are highly non-trivial. Moreover, a full
understanding of the Gribov problem necessary for eliminating the
gauge degrees of freedom is beyond present days knowledge.  Hence a
non-perturbative treatment of the divergencies in the fermion sector
can only be a first, though probably very important, step towards a
deeper understanding of the gauge theories we are ultimately
interested in, e.g\ QCD$_{3+1}$.
\app
\section*{Appendix A}
In this Appendix we prove the explicit formula
\Ref{CC2} for the 2-cocycle $\chi(U,V;F)$, valid for all
$F\in\Grhe{1}$ and $U,V\in\Ghe{1}$ in some neighborhood of the
identity.

{}From the results of Ruijsenaars \cite{R1} one learns that for
$U\in\GhF{1}\cap\UhF$, the implementer of the Bogoliubov transformation
$\alpha_U$ in $\Pi_F$ can be written as \footnote{
the prime is used to indicate that the phase convention
for these implementers differs from the one leading to
the $F$-independent 2-cocycle \Ref{CC2}
}
\eq
\label{A.1}
\hGamp{U;F} = E(U;F)N'(U;F)
\eqend
with
\eqa
\label{A.2}
E(U;F) = \ee{Z_{+-}(U;F)a^*a^*}
\FGam{Z_{++}(U;F)\oplus Z_{--}(U;F)^t}\nonu
\times\ee{-Z_{-+}(U;F)aa}
\eqaend
($(\cdot)^t\equiv C_F(\cdot)^*C_F$)
an operator on $\cF(h)$ satisfying
\eq
\label{A.3}
\Produ{\Om}{E(U;F)\Om} = 1
\eqend
($\Produ{\cdot}{\cdot}$ is the inner product in $\cF(h)$),
and
\eq
\label{A.4}
N'(U;F)=  \det{1+(Z_{+-}(U;F))^*Z_{+-}(U;F)}^{-1/2}
\eqend
a normalization constant; here we introduced
($P_\pm=\f{1}{2}(1\pm F)$)
\eqa
\label{A.5}
Z_{++}(U;F) &=& P_+UP_+ - P_+UP_-[P_-UP_-]^{-1}P_-UP_+ \nonu
Z_{+-}(U;F) &=& P_+UP_-[P_-UP_-]^{-1}\nonu
Z_{-+}(U;F) &=& -[P_-UP_-]^{-1}P_-UP_+\nonu
Z_{--}(U;F) &=& [P_-UP_-]^{-1}
\eqaend
and used the notation
\eqa
\label{Aaa}
A_{+-}a^*a^* &=& \sum_{n,m=1}^{\infty} \produ{f_n^{+}}{A_{+-}f_m^{-}}
a^*(f_n^{+}) a^*(C_F f_n^{-})\nonu
A_{-+}aa &=& \sum_{n,m=1}^{\infty} \produ{f_n^{-}}{A_{-+}f_m^{+}}
a(C_F f_n^{-}) a(f_n^{+})
\eqaend
well-defined for Hilbert-Schmidt operators $A_{\pm\mp}=P_\pm A_{\pm\mp}P_\mp$
with $\{f_n^{\pm}\}_{n=1}^\infty$ some complete, orthonormal
bases in $P_\pm h$\footnote{
see Ref.\ \cite{BT4}; note that $Z_{\pm\mp}(U;F)\in\I{2}$ (cf.\ Appendix B)
}
(see Ref.\ \cite{R2}, eqs.\ (2.15) and (3.8) and \cite{R4}, Appendix D).

Let $U,V\in\GhF{1}\cap\UhF$. From
\eq
\label{A.6}
\hGamp{U;F}\mult\hGamp{V;F} = \chi'(U,V;F) \hGamp{UV;F}
\eqend
and \Ref{A.1}, \Ref{A.3} one can deduce that
\eq
\label{A.7}
\chi'(U,V;F) = \f{N'(V;F^U)N'(U;F)}{N'(UV;F)}\cE(U,V;F)
\eqend
with
\eqa
\label{A.8}
\cE(U,V;F) = \Produ{\Om}{\FGam{U}E(V;F^U)\FGam{U}^*E(U;F)\Om} \nonu
= \Produ{\Om}{\FGam{U}\ee{-Z_{-+}(V;F^U)aa}\FGam{U}^*
\ee{Z_{+-}(U;F)a^*a^*}\Om}
\eqaend
(we used $\FGam{\cdots}\Om = \ee{(\cdots)aa}\Om = \Om$).

By \Ref{A.5}, \Ref{Aaa} and \Ref{2.4}
\[
\FGam{U}(Z_{-+}(V;F^U)aa)\FGam{U}^*= (Z_{-+}^U(V;F)aa)
\]
with
\[
Z_{-+}^U(V;F) = UZ_{-+}(V;F^U)U^* =
Z_{-+}(UVU^*;F),
\]
hence with the formula
\[
\Produ{\Om}{\ee{A_{-+}aa}\ee{B_{+-}a^*a^*}\Om} =
\det{1+ A_{-+}B_{+-}}
\]
(see Theorem 3.2 in Ref.\ \cite{R3}) one obtains
\eq
\label{A.9a}
\cE(U,V;F) = \det{1+ [^UV_{--}]^{-1}(^UV_{-+})U_{+-}[U_{--}]^{-1} }
\eqend
where we introduced the notation
\eq
\label{A.10}
A_{\epsilon\epsilon'}\equiv P_\epsilon AP_{\epsilon'}\quad \forall
\epsilon,\epsilon'\in\{+,-\}
\eqend
and
\eq
\label{A.11}
^UV\equiv UVU^{*}.
\eqend
Using $(^UV_{-+}U_{+-})=(^UVU)_{--}-(^UV_{--}U_{--})$ and introducing the
determinant
\eq
\detF{(\cdots)_{--}}\equiv \det{P_+ +(\cdots)_{--}}
\eqend
in $P_-h$, this can be written as
\eq
\label{A.9}
\cE(U,V;F) = \detF{[^UV_{--}]^{-1}(^UVU)_{--}[U_{--}]^{-1}}.
\eqend
Similarly, we obtain for \Ref{A.4}
\eqa
\label{Nprime}
N'(U;F)& =& \left(\det{1+[U^*_{--}]^{-1}(U^*)_{-+}U_{+-}[U_{--}]^{-1}}
\right)^{-1/2} \nonu &=&
\left(\detF{[U^*_{--}]^{-1}[U_{--}]^{-1} } \right)^{-1/2}.
\eqaend
As $^UVU=UV$ and
\[
\det{U^*\cdots U}=\det{\cdots},
\]
one has $N'(UV;F)=N'(^UVU;F)$ and $N'(V;F^U)=N'(^UV;F)$, and
\Ref{A.7} results in
\eq
\label{A.12}
\chi'(U,V;F) = \left(\f{\detF{[^UV_{--}]^{-1}(^UVU)_{--}[U_{--}]^{-1} }}%
{\detF{[^UV_{--}]^{-1}(^UVU)_{--}[U_{--}]^{-1}}^*}\right)^{1/2}
\eqend
where we used repeatedly some basic properties of determinants.
It is easy to check that the determinants in \Ref{A.12} exist for
$U,V\in\GhF{1}\cap\UhF$.\footnote{
use
\[
[^UV_{--}]^{-1}(^UVU)_{--}[U_{--}]^{-1} =
P_- + [^UV_{--}]^{-1} (^UV_{-+}U_{+-})[U_{--}]^{-1}
\]
and the fact that $U_{\pm\mp}\in\I{2}$ for $U\in\GhF{1}$
(see Appendix B)
}
Moreover, if $U,V$ are even in $\GhF{0}\cap\UheF$, we can write
\Ref{A.12} as
\eq
\label{LLL}
\chi'(U,V;F)= \f{\beta_1'(^UV;F)\beta_1'(U;F)}{\beta_1'(^UVU;F)}
\eqend
with
\eq
\beta_1'(U;F)=\left(\f{\detF{U_{--}^*}}{\detF{U_{--}}}\right)^{1/2}
=\left(\f{\det{P_+ +P_-U^*P_-}}{\det{P_+ +P_-UP_-}}\right)^{1/2}.
\eqend
As $^UVU=UV$ and $\beta_1'(^UV;F)=\beta_1'(V;F^U)$, we can write
\Ref{LLL} locally as coboundary:
$\chi'=\del\beta_1'$ ($\del$ cf.\ \Ref{b.8}; note that $\beta_1'$
obeys \Ref{111}).

Thus for $U,V\in\Ghe{0}\cap\UheF$, we can transform $\chi'$ to a
$F$-independent expression by multiplying it with the 2-coboundary
$\del\beta$ where locally $\beta(U;F)=\beta_1'(U;\eps)/\beta_1'(U;F)$,
i.e.\
\[
\beta(U;F) = \left(\f{\det{1+a(U;F)}}{\det{1+a(U;F)}^*}
\right)^{1/2}
\]
with
\eq
\label{defa}
a(U;F) = (P_++P_-UP_-)(P_+^0+[P_-^0UP_-^0]^{-1}) -1 \: .
\eqend
It is convenient to write this as
\eq
\label{A.14}
\beta(U;F)= \left(
\f{\detz{1+a(U;F)}\exp{(\tra{P_+^0a(U;F)P_+^0 +
P_-^0a(U;F)P_-^0})}}{\detz{1+a(U;F)}^*\exp{(\tra{P_+^0a(U;F)P_+^0 +
P_-^0a(U;F)P_-^0}^*)}}\right)^{1/2}
\eqend
(we used $\det{1+a}=\detz{1+a}\exp{(\tra{a_{++}+a_{--}})}$ which
follows from $\tra{a_{\pm\mp}}=0$ and the definition of
$\detz{\cdot}$).  Using the Lemma in Appendix B one can easily prove
that for $U\in\Ghe{1}\cap\UheF$, $F\in\Grhe{1}$, one has
$a(U;F)\in\I{2}$, hence $\beta(U;F)$ \Ref{A.14} exists. It follows
that $\del\beta$ is locally a 2-coboundary of $\Ghe{1}$, and the
implementers
\eq
\label{A.16}
\hGam{U;F}\equiv \hGamp{U;F}\beta(U;F)
\eqend
obey \Ref{mult} with the 2-cocycle $\chi=\del\beta\chi'$ which is
(locally) $F$-independent and equal to \Ref{CC2}. Note that
\eq
\label{A.18}
\hGam{U;F}= E(U;F)N(U;F)
\eqend
with
\eq
\label{A.19}
N(U;F)=N'(U;F)\beta(U;F)
\eqend
equal to \Ref{N}.
\appende
\app

\section*{Appendix B}
In this Appendix we summarize the basic properties of operators
in $\Ghe{p}$, $\ghe{p}$ and $\Grhe{p}$, $p\in\N$.

For $\eps$ a grading operator, $P_\pm^0=\f{1}{2}(1\pm \eps)$
are orthogonal projections on $h$, and we introduce the
notation
\eq
\label{B.1}
A_{\epsilon\epsilon'}\equiv P^0_\epsilon AP^0_{\epsilon'}\quad \forall
\epsilon,\epsilon'\in\{+,-\}
\eqend
for all linear operators $A$ on $h$;
moreover, it is sometimes convenient to use the following
2$\times$2-matrix notation:
\eq
\label{matrix}
A=\left(\bma{cc} A_{++}&A_{+-}\\ A_{-+}& A_{--}\ema\right),
\quad \mbox{i.e.\ } \eps=\left(\bma{cc}1&0\\0&-1\ema\right).
\eqend

\subsubsection*{Lemma:}
\begin{itemize}
\item[$(i)$] A unitary operator $U$ on $h$ is in $\Ghe{p}$ if and only if
\[
U_{+-},U_{-+}\in\I{2p}.
\]
\item[$(ii)$] A bounded, self-adjoint operator $u$ on $h$ is in
$\ghe{p}$ if and only if
\[
u_{+-},u_{-+}\in\I{2p}.
\]
\item[$(iii)$] A grading operator $F$ on $h$ is in
$\Grhe{p}$ if and only if
\[
F_{+-},F_{-+}\in\I{2p}, \quad (F-\eps)_{++},(F-\eps)_{--}\in\I{p}.
\]
\end{itemize}
{\bf Proof:} By using the matrix notation \Ref{matrix},
$(i)$ and $(ii)$ immediately follow from
\eq
\label{com}
\ccr{\eps}{A} = 2\left(\bma{cc}0&A_{-+}\\ -A_{-+}&0\ema\right).
\eqend
To prove $(iii)$, we note that by definition, $F\in\Grhe{p}$ if and
only if $(F-\eps)\in\I{2p}$. This is equivalent to
\[
(F-\eps)_{\pm\pm}, F_{\pm\mp}\in\I{2p}.
\]
But as $F^2=1$,
\nonueqa
(F-\eps)^*_{\pm\pm}(F-\eps)_{\pm\pm} = F_{\pm\pm}F_{\pm\pm}\mp
2F_{\pm\pm}+ P_\pm^0\nonu =P_\pm^0 -F_{\pm\mp}F_{\mp\pm}\mp
2F_{\pm\pm}+P^0_\pm = \mp
2(F-\eps)_{\pm\pm}-(F_{\mp\pm})^*F_{\mp\pm},
\nonueqaend
hence for $F_{\pm\mp}\in\I{2p}$, $(F-\eps)_{\pm\pm}\in\I{2p}$ if and
only if it is in $\I{p}$.
\appende
\app

\section*{Appendix C}
In this Appendix we evaluate the Kac-Peterson cocycle \Ref{KP}.

Assuming that $U,V\in\Ghe{0}$ and close to the identity, the
2-cocycle \Ref{CC2} can be written as 2-coboundary: $\chi=\del
\beta_1$, with $\beta_1$ \Ref{CB2}.  Thus for $u,v\in\ghe{0}$, we
have by our general discussion in Section \ref{sec3.2} $c_1=-\dd
b_1$ with
\[
b_1(u;F)=\f{\dd}{\ii\dd t}\beta_1(\ee{\ii tu};F)\restr{t=0}
\]
and $\dd$ defined in \Ref{cob}. As
\[
P_-^0\ee{\ii tu}P_-^0-P_-^0= \ii tu_{--} + \OO(t^2)
\]
(using the notation introduced in the Appendix B), we obtain
\[
b_1(u;F)= -\tra{u_{--}}.
\]
Therefore
\eqa
\label{KPC}
c(u,v;F)= -\dd b_1(u,v;F)= -b_1(\ccr{u}{v};F) = \nonu
\tra{(uv)_{--}-(vu)_{--}} = \tra{u_{-+}v_{+-}-v_{-+}u_{+-}}
\eqaend
by the cyclicity of the trace. This is finite for all $u,v\in\Ghe{1}$,
and identical with \Ref{KP} (as is easily proved by using
the matrix notation \Ref{matrix}).

\appende
\app

\section*{Appendix D}
In this Appendix we prove that $\chi_2(U,V;F)$ \Ref{chip} ($p=2$) with
$\beta_2(U;F)$ \Ref{CB4} is of the form \Ref{regu}, hence can be
extended to all $F\in\Grhe{2}$ and $U,V\in\Ghe{2}\cap\UheF$.

For $U,V\in\Ghe{1}\cap\UheF$, $F\in\Grhe{1}$, the 2-cocycle \Ref{CC2} can be
written as
\eq
\chi(U,V;F)=\left(\f{X(U,V;F)}{X(U,V;F)^*}\right)^{1/2}
\eqend
with (we use the notation from Appendix B)
\eq
\label{C.2}
X(U,V;F) =\det{1+[U_{--}]^{-1}U_{-+}V_{+-}[V_{--}]^{-1}}.
\eqend
Similarly, the 1-cochain \Ref{CB4} is
\eq
\label{C.3}
\beta_2(U;F)=\left(\f{B_2(U;F)}{B_2(U;F)^*}\right)^{1/2}
\eqend
with
\eq
B_2(U;F) = \exp{(\f{1}{2}\tra{F_{-+}U_{+-}U_{--}^*})}.
\eqend
As \Ref{C.2} is equivalent to
\nonueqa
X(U,V;F) = \exp{(\tra{[U_{--}]^{-1}U_{-+}V_{+-}[V_{--}]^{-1}}}
\nonu \times \detz{1+[U_{--}]^{-1}U_{-+}V_{+-}[V_{--}]^{-1}},
\nonueqaend
all we have to show is that
\[
\del B_2(U,V;F)\exp{(\tra{[U_{--}]^{-1}U_{-+}V_{+-}[V_{--}]^{-1}})}
\equiv\exp{(\tra{Y_2(U,V;F)})}
\]
allows for a continuous extension to $U,V\in\Ghe{2}\cap\UheF$,
$F\in\Grhe{2}$.

We have
\eqa
Y_2(U,V;F)= [U_{--}]^{-1}U_{-+}V_{+-}[V_{--}]^{-1}
+\f{1}{2}F_{-+}U_{+-}U_{--}^* \nonu
 + \f{1}{2}(F^U)_{-+}V_{+-}V_{--}^* -
\f{1}{2}F_{-+}(UV)_{+-}(V^*U^*)_{--} \: .
\eqaend
The relation $[U_{--}]^{-1}U_{--} = (U^*U)_{--} = U^*_{--}U_{--} +
(U^*)_{-+}U_{+-}$ implies
\[
[U_{--}]^{-1} = U^*_{-} + (U^*)_{-+}U_{+-}[U_{--}]^{-1};
\]
using also $(UV)_{+-}=U_{+-}V_{--}+U_{++}V_{+-}$ and similar relation
for $(F^U)_{-+}=(U^*FU)_{-+}$ and $(V^*U^*)_{--}$ we obtain (the
symbol ``$\sim$'' below means ``equal up to terms which are obviously
trace-class for all $U,V\in\Ghe{2}$, $F\in\Grhe{2}$'')
\[
[U_{--}]^{-1}U_{-+}V_{+-}[V_{--}]^{-1}\sim
U_{--}^*U_{-+}V_{+-}V_{--}^*,
\]
\nonueqa
(F^U)_{-+}V_{+-}V_{--}^*\sim U_{--}^*F_{--}U_{-+}V_{+-}V_{--}^*
+U_{--}^*F_{-+}U_{++}V_{+-}V_{--}^* \\
+U_{-+}^*F_{++}U_{++}V_{+-}V^*_{--}
\sim -2U^*_{--}U_{-+}V_{+-}V^*_{--}
+U^*_{--}F_{-+}U_{++}V_{+-}V^*_{--},
\nonueqaend
(we used $U^*_{-+}U_{++}=-U^*_{--}U_{-+}$),
\nonueqa
F_{-+}(UV)_{+-}(V^*U^*)_{--}\sim F_{-+}U_{++}V_{+-}V^*_{--}U^*_{--}
+ F_{-+}U_{+-}V_{--}V^*_{--}U^*_{--}\\
\sim F_{-+}U_{++}V_{+-}V^*_{--}U^*_{--}
+F_{-+}U_{+-}U^*_{--},
\nonueqaend
hence
\[
\tra{Y_2(U,V;F)}\sim\tra{\f{1}{2}U^*_{--}F_{-+}U_{++}V_{+-}V^*_{--}
-\f{1}{2}F_{-+}U_{++}V_{+-}V^*_{--}U^*_{--}} = 0
\]
due to the cyclicity of the trace.

\appende
\app

\section*{Appendix E}
In this Appendix we evaluate the Mickelsson-Rajeev cocycle
\cite{MR}.
{}From our general discussion in Section \ref{sec3.2} we have
\[
c_2(u,v;F)= c(u,v;F)-\dd b_2(u,v;F)
\]
with $c_1$ the Kac-Peterson cocycle \Ref{KPC} and
\[
b_2(u;F)= \f{\dd}{\ii\dd t} \beta_2(\ee{\ii tu};F)\restr{t=0};
\]
with $\beta_2$ \Ref{CB4} we obtain (we use the
notation introduced in the Appendix B)
\[
b_2(u;F)=\f{1}{4}\tra{u_{-+}F_{+-}+u_{+-}F_{-+}} =
-\f{1}{16}\tra{\ccr{u}{\eps}\ccr{F}{\eps}},
\]
and
\[
\dd b_2(u,v;F)=b_2(\ccr{u}{v};F)-b_2(v;\ccr{F}{u})+b_2(u;\ccr{F}{v})
\]
which gives (after a lengthy but straightforward calculation)
\[
\dd b_2(u,v;F) = \f{1}{2}
\tra{F_{++}(v_{+-}u_{-+}-u_{+-}v_{-+})
-F_{--}(u_{-+}v_{+-}-v_{-+}u_{+-})}
\]
hence
\nonueqa
c_2(u,v;F)= \f{1}{2}{\rm tr}((F-\eps)_{--}(u_{-+}v_{+-}-v_{-+}u_{+-})
\\ -(F-\eps)_{++}(v_{+-}u_{-+}-u_{+-}v_{-+}))
\nonueqaend
identical with \Ref{MR}.
\appende

\bc
\section*{Acknowledgement}
\ec
I would like to thank A. L. Carey, H.~Grosse, G.~Kelnhofer, J.
Mickelsson, S.~N.~M.~Ruijsenaars, M. Salmhofer, and G. Semenoff for
valuable discussions and comments. Furthermore I am indebted to M.
Salmhofer for carefully reading the manuscript and to S. N. M.
Ruijsenaars for pointing out some errors in the first version of this
paper.  It is a pleasure to thank H. Grosse for his encouragement and
his interest in my work.  Financial support of the
``Bundeswirtschaftkammer'' of Austria is appreciated.
\vspace*{1cm}

\end{document}